\documentclass[lettersize,journal]{IEEEtran}
\usepackage{amsmath,amsfonts}
\usepackage{algorithm}
\usepackage{array}
\usepackage[caption=false,font=normalsize,labelfont=sf,textfont=sf]{subfig}
\usepackage{multirow}
\usepackage{textcomp}
\usepackage{stfloats}
\usepackage{url}
\usepackage{colortbl}
\usepackage{verbatim}
\usepackage{graphicx}
\usepackage{soul}
\soulregister\cite7 
\soulregister\citep7 
\soulregister\citet7 
\soulregister\ref7 
\soulregister\pageref7 
\usepackage{multirow}
\usepackage[table,xcdraw]{xcolor}
\usepackage{xcolor}
\usepackage{cite}
\floatname{algorithm}{Algorithm}
\usepackage{algpseudocode}
\usepackage{amsmath}
\usepackage[switch]{lineno}

\begin{document}
\title{Real-Time AIoT for AAV Antenna Interference Detection via Edge–Cloud Collaboration}
\author{Jun Dong,~\IEEEmembership{Student Member,~IEEE,} Jintao Cheng, Jin Wu,~ \IEEEmembership{Member,~IEEE,} Chengxi Zhang, Shunyi Zhao,~ \IEEEmembership{Senior Member,~IEEE,} Xiaoyu Tang,~ \IEEEmembership{Member,~IEEE}

\thanks{This research was supported by the National Natural Science Foundation of China under grants 62001173 and 62233013, the Project of Special Funds for the Cultivation of Guangdong College Students  Scientific and Technological Innovation (Climbing Program  Special Funds) under grants pdjh2022a0131 and pdjh2023b0141, the Science and Technology Commission of Shanghai Municipal under grant 22511104500, the Fundamental Research Funds for the Central Universities, and the Xiaomi Young Talents Program. (Corresponding author: Xiaoyu Tang).

Jun Dong is with the School of Data Science and Engineering, and Xingzhi College, South China Normal University, Shanwei, 516600, China (e-mail: 20228132044@m.scnu.edu.cn).

Jintao Cheng is with the School of Physics, South China Normal University, Guangzhou, 510006, China(e-mail: chengjt\_alex@outlook.com).

Jin Wu is with the Department of Electronic and Computer Engineering, Hong Kong University of Science and Technology, Hong Kong, China(e-mail: jin\_wu\_uestc@hotmail.com).

Chengxi Zhang and Shunyi Zhao are with the Key Laboratory
of Advanced Control for Light Industry Processes, Ministry of Education,
Jiangnan University, Wuxi 214122, China (e-mail: dongfangxy@163.com; shunyi@jiangnan.edu.cn).

Xiaoyu Tang is with the School of Electronics and Information Engineering,
and Xingzhi College, South China Normal University, Shanwei, 516600, (email: tangxy@scnu.edu.cn)
}
}

\markboth{Journal of \LaTeX\ Class Files,~Vol.~14, No.~8, August~2021}%
{Shell \MakeLowercase{\textit{et al.}}: A Sample Article Using IEEEtran.cls for IEEE Journals}

\IEEEpubid{\begin{minipage}{\textwidth}\ \\[25pt] \centering
		Copyright \copyright 20xx IEEE. Personal use of this material is permitted.
		However, permission to use this material for any other purposes must \\ be obtained
		from the IEEE by sending an email to pubs-permissions@ieee.org.
\end{minipage}}

\maketitle

\begin{abstract}
In the fifth-generation (5G) era, eliminating communication interference sources is crucial for maintaining network performance. Interference often originates from unauthorized or malfunctioning antennas, and radio monitoring agencies must address numerous sources of such antennas annually. Unmanned aerial vehicles (UAVs) can improve inspection efficiency. However, the data transmission delay in the existing cloud-only (CO) artificial intelligence (AI) mode fails to meet the low latency requirements for real-time performance. Therefore, we propose a computer vision-based AI of Things (AIoT) system to detect antenna interference sources for UAVs. The system adopts an optimized edge-cloud collaboration (ECC+) mode, combining a keyframe selection algorithm (KSA), focusing on reducing end-to-end latency (E2EL) and ensuring reliable data transmission, which aligns with the core principles of ultra-reliable low-latency communication (URLLC). At the core of our approach is an end-to-end antenna localization scheme based on the tracking-by-detection (TBD) paradigm, including a detector (EdgeAnt) and a tracker (AntSort). EdgeAnt achieves state-of-the-art (SOTA) performance with a mean average precision (mAP) of 42.1\% on our custom antenna interference source dataset, requiring only 3 million parameters and 14.7 GFLOPs. On the COCO dataset, EdgeAnt achieves 38.9\% mAP with 5.4 GFLOPs. We deployed EdgeAnt on Jetson Xavier NX (TRT) and Raspberry Pi 4B (NCNN), achieving real-time inference speeds of 21.1 (1088) and 4.8 (640) frames per second (FPS), respectively. Compared with CO mode, the ECC+ mode reduces E2EL by 88.9\%, increases accuracy by 28.2\%. Additionally, the system offers excellent scalability for coordinated multiple UAVs inspections. The detector code is publicly available at https://github.com/SCNU-RISLAB/EdgeAnt.

\end{abstract}

\section{Introduction}
\IEEEPARstart{T}{HE} ultrahigh bandwidth and controllable latency of fifth-generation (5G) technology are driving the rapid adoption of edge computing and Internet of Things (IoT) devices\cite{ref1}. These interconnected devices enable real-time data collection and processing, forming the backbone of modern intelligent services\cite{ref2}. Residents' unauthorized installation of antennas amplifies and retransmits repeater signals without regulation, causing spectrum congestion in the same geographical areas and frequency bands, thereby undermining the stability of legitimate base station communication systems\cite{ref3}. By detecting these antennas, the locations of such interference devices can be quickly identified. During routine inspections, spectrum analyzers can only approximate the locations of interference sources, necessitating onsite searches to remove them. Wang Rui's\cite{ref4} work on radio interference source positioning via handheld monitoring receivers introduced new radio monitoring methods. However, interference sources often hide in hard-to-observe locations, complicating daily troubleshooting efforts.

Cloud video surveillance (CVS) is gaining attention because of advancements in the IoT and computer vision technology\cite{ref5}. Utilizing unmanned aerial vehicles (UAVs) equipped with cameras to patrol antennas in hard-to-reach locations can significantly improve work efficiency, enabling large-scale interference source detection. CVS systems are typically implemented via cloud computing to accommodate high-complexity and high-precision neural network models\cite{ref6}. This computational mode, in which video data are uniformly processed by servers, requires high bandwidths when transmitting video streams, resulting in severe data transmission delays and transmission energy consumption levels. Fog computing technology\cite{ref7} adopts a distributed computing approach, distributing computational, communication, control, and storage resources and services to devices and systems that are close to the edge. Although this strategy operates closer to the edge than cloud computing does, relying only on fog node processing is insufficient for addressing the high latency caused by video streaming transmissions. Edge computing introduces local computation, which can effectively reduce latency caused by communication while ensuring privacy and data security\cite{ref8}. Fig. 1 shows IoT solutions implemented in three different computational modes.

For low-latency sensitive IoT applications like UAV inspections, the high latency caused by cloud-only (CO) mode video streaming significantly affects positioning accuracy\cite{ref9}. The limited computing resources in an edge-only (EO) mode are insufficient to support model retraining on continuously expanding datasets. Due to the inherent limitations of a single mode, edge computing is increasingly being combined with cloud computing to fully leverage its communication, storage, and computational capabilities, forming an edge-cloud collaboration mode (ECC) \cite{ref10}. Some well-trained AI models on cloud servers are deployed on the edge side, effectively solving the problems regarding the time-consuming CO mode and the isolation of a single edge computing server through distributed and reliable computing.

\begin{figure}[]
\centering
\includegraphics[width=3.5in]{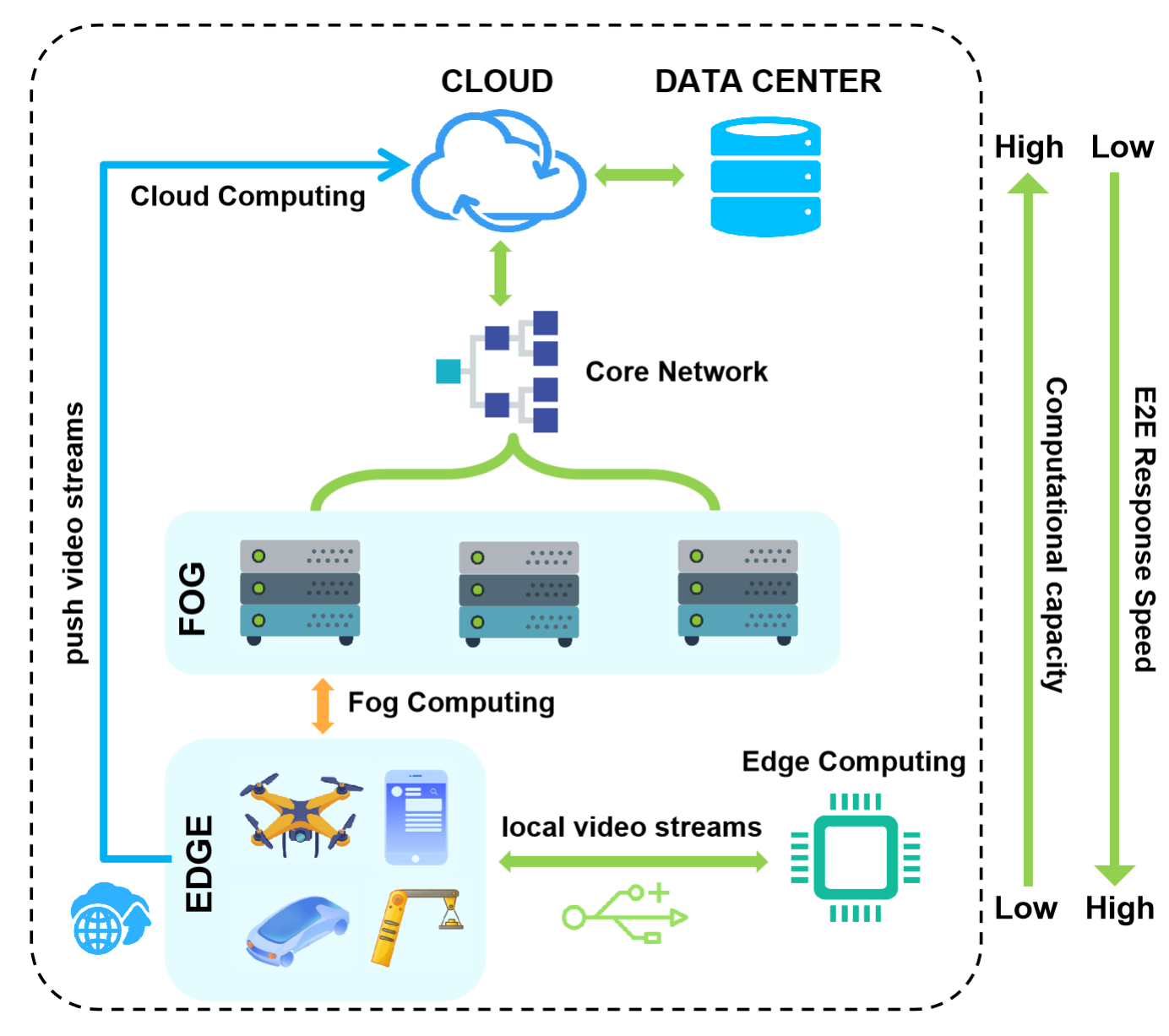}
\caption{
Three different computing modes that are common to AIoT applications include cloud computing (blue path), fog computing (orange path), and edge computing (green path).
}
\label{fig1}
\end{figure}

Deep convolutional neural networks (DCNNs) have been widely studied because of their fast, scalable, and end-to-end learning frameworks. The most representative approach is the You Only Look Once (YOLO)\cite{ref11} series of object detection networks. However, onboard computing power is still the most prominent factor limiting the ability to perform real-time embedded computer vision processing in UAVs\cite{ref12}. The complex background also poses challenges for detection. Fig. 2 illustrates three familiar sources of antenna interference. The Yagi antenna is an end-fire antenna composed of parallel elements, and its hollow structure makes it difficult for detectors to define the antenna's edges accurately. The Patch antenna and Plate log antenna are usually light-colored and rectangular. In practical scenarios, antennas are usually placed on high-rise buildings or balconies with fences. Due to confusion and color similarities between the target and the background, this leads to false positives and false negatives. Low-resolution images fail to capture sufficient antenna features. In contrast, high-resolution images lead to increased system memory usage and energy consumption, significantly reducing real-time inference speed and the operational time of UAVs\cite{ref13}. Unfortunately, the existing lightweight detectors fail to balance accuracy and efficiency in practical interference source inspection tasks.

\begin{figure}[]
\centering
\includegraphics[width=3.5in, height=3.6in]{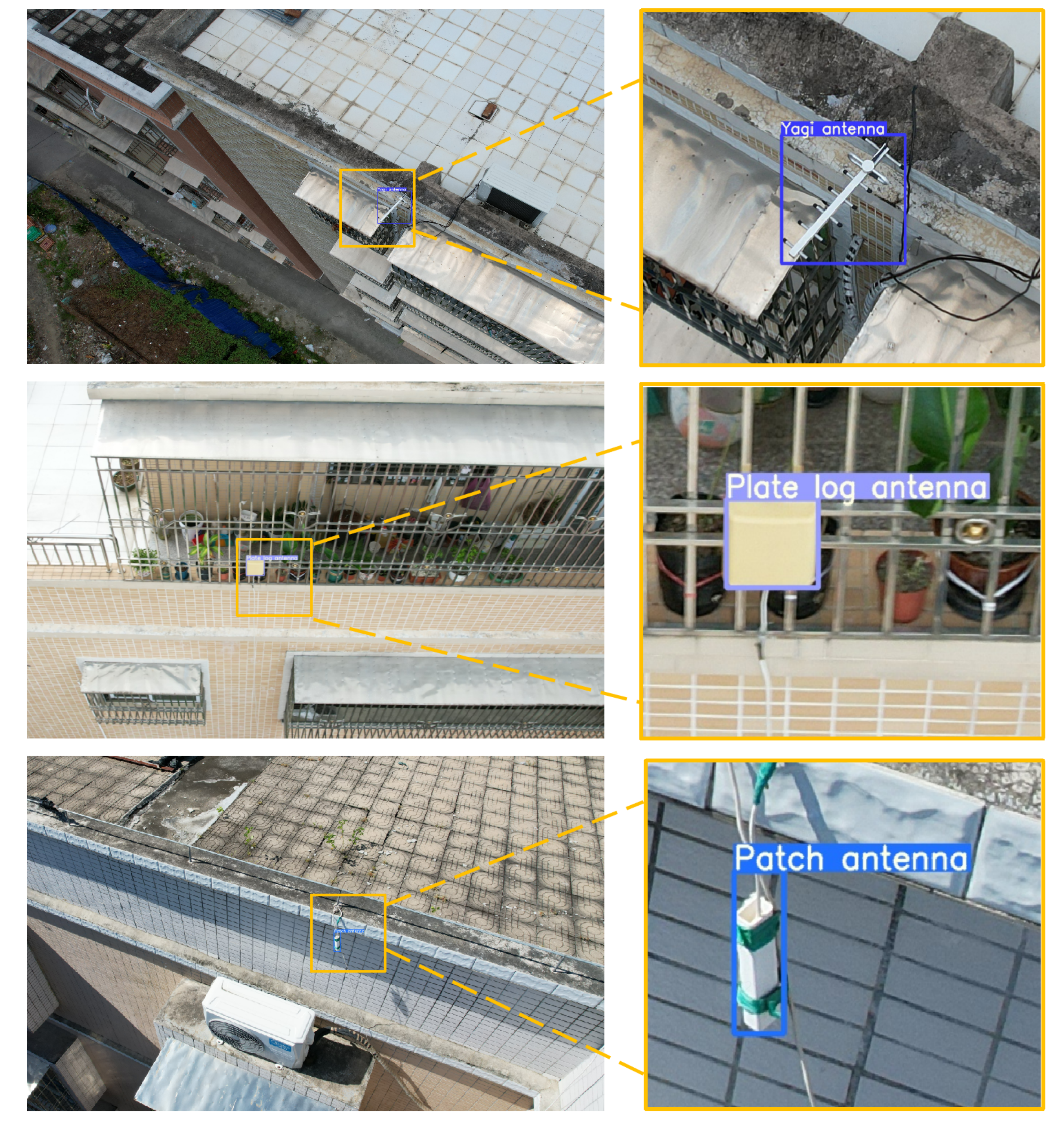}
\caption{Three common types of antenna interference sources in actual aerial images: Yagi antenna, Plate log antenna, and Patch antenna.}
\label{fig2}
\end{figure}

Multiobject tracking (MOT) aims to detect and estimate the spatiotemporal trajectories of multiple objects in a video stream. MOT is a fundamental problem in many application systems, such as video surveillance\cite{ref14}. Benefiting from high-performance detectors, the tracking-by-detection (TBD) paradigm has gained popularity because of its excellent performance. This paradigm is also applicable to the task of inspecting antenna interference sources. However, the recent research on trackers has focused primarily on the use of various reidentification (ReID) models to achieve improved MOT performance\cite{ref15}, aiming for high tracking accuracy. These approaches were designed mainly for crowd-tracking scenarios, and they are not suitable for the direct practical application of inspecting antennas.

The existing methods cannot be directly applied to UAV interference source inspection tasks, and they need help due to their low system mode efficiency, poor detector performance, and unsuitable trackers. Therefore, we design an artificial intelligence of things (AIoT) system based on the ECC+ mode. The inference results are selectively uploaded to the cloud server through the keyframe selection algorithm (KSA) compared to the ECC mode. At the edge layer, we propose an E2E antenna interference source localization scheme based on the TBD paradigm, which is encapsulated and deployed on a UAV with an edge computing device. The TBD solution consists of EdgeAnt, a lightweight detector, and AntSort, a tracker dedicated to interference source inspection. In summary, the main contributions of this paper are as follows

\begin{enumerate}
\item{An efficient real-time AIoT antenna interference source detection system based on ECC+ is proposed. We provide more precise definitions for the tasks implemented at the edge and in the cloud. When the proposed KSA is utilized, the system achieves more streamlined and efficient communication between the edge and the cloud. Furthermore, we developed a model for coordinated inspection path planning with multiple UAVs and evaluated the system's scalability.}
\item{A new baseline detector, EdgeAnt, is developed. We design a lightweight backbone network called the lightweight hierarchical geometric network (LHGNet) and a neck network called the heterogeneous bidirectional feature pyramid network (HetBiFPN). We also introduce a small object enhancement layer (EL) composed of a two-segment residual block (TSRBlock) to enhance the understanding of small objects. After integrating with the improved tracker AntSort, the system balances end-to-end latency (E2EL) and inspection accuracy.}
\item{A performance evaluation is conducted on the detector. EdgeAnt achieved a SOTA performance of 42.1\% mAP on the antenna interference dataset with 3.0 million parameters and 14.7 GFlops. It achieves the fastest real-time inference speed on edge computing devices, reaching 21.1 frames per second (FPS) on Jetson Xavier NX (TRT) and 4.8 FPS on Raspberry Pi 4B (NCNN). It is equally competitive on the COCO dataset.}
\item{A performance evaluation is conducted on the system. Compared with those of the solutions implemented in the cloud mode and the ECC mode, the E2EL of the system in the ECC+ mode is reduced by 88.9\% and 62.5\%, respectively. Moreover, we evaluated input video streams of various resolutions and scenarios with fluctuating bandwidth, showcasing the system's robust stability. Simulation results show that coordinating multiple UAVs can achieve lower communication latency and inspection time in practical applications.}
\end{enumerate}

The rest of the paper is organized as follows. Section II reviews the related work. Section III introduces the composition and workflow of the AIoT interference source detection system in the ECC+ mode, including the KSA and a multiple UAVs coordinated inspection path modeling. It also details the localization schemes of the TBD paradigm with the EdgeAnt detector and AntSort tracker. Section IV conducts thorough experimental testing on the detector and the system separately. Section V concludes with an overview of the strengths, weaknesses, and future improvement directions of the proposed system.

\section{Related work}
In this section, we first introduce intelligent video surveillance technology and its application and development history concerning edge computing in Section A. We then review the object detection methods developed for UAV IoT applications and video surveillance systems in Section B.
\subsection{Video Surveillance and Edge Computing}

The rapid development of DCNNs has led to methods such as FGFA\cite{ref16} and MEGA\cite{ref17}, which utilize inter-frame correlations to achieve accurate offline video object detection. With the widespread adoption of online video surveillance\cite{ref18}, Zhu et al.\cite{ref19} proposed Deep Feature Flow (DFF), which runs complex convolutional networks only on sparse key frames and propagates their deep feature maps to other frames through a flow field to improve recognition speed further. However, this method can only run on the cloud, and the limitations of video data transmission in edge environments restrict the system's applicability in real-time scenarios.

Researchers worldwide have made significant efforts to achieve always-online AI at the near end of the IoT\cite{ref20}. Compared to the CO mode, deploying pure inference AI at the edge can achieve lower-latency IoT services\cite{ref21}. Yi et al.\cite{ref22} proposed a lightweight crowd counting network (LEDCrowdNet), and the algorithm was successfully deployed on two edge computing platforms: the NVIDIA Jetson Xavier NX and the Coral Edge TPU. Vikas Goyal et al.\cite{ref23} proposed an edge IoT-based model to monitor and detect anomalies in the internal environment of a farm, and a Raspberry Pi 4 device with limited computational resources was used to implement the application. Liu et al.\cite{ref24} deployed a keyframe algorithm based on CNN and AT-YOLO on the Raspberry Pi 4B, achieving an inference speed of 13.69 FPS. 

For detection and tracking tasks, \cite{ref25} proposed $EC^2$Detect, an ECC method where target detection occurs on the cloud according to adaptively selected keyframes, while edge devices handle lightweight tracking. This system does not eliminate the dependence on cloud computing during the inference process and still inevitably incurs delays due to uploading most frames. In contrast, our ECC+ mode-based system delegates all detection and tracking tasks to edge servers via the KSA.

\subsection{UAV Object Detection}
UAV platforms in the IoT need to sense their environments, understand scenarios and react accordingly\cite{ref26}. Advanced and computationally expensive algorithms cannot be directly applied to embedded devices. Furthermore, detecting small objects in UAV images is challenging because of the scale variations exhibited by targets. To address the problem of insufficient contextual information for small targets, Du et al.\cite{ref27} proposed a global context-enhanced adaptive sparse convolutional network (CEASC). Many researchers are also working to adapt YOLOv8 for use in UAV aerial photography scenarios\cite{ref28,ref29}. Xiong et al.\cite{ref30} introduced the AS-YOLOv5 algorithm, which features adaptive fusion and an improved attention mechanism, achieving good performance on public datasets. Mao et al.\cite{ref31} employed a cost-effective split-and-shuffle operation to reduce model inference memory and computational costs. Zheng et al.\cite{ref32} proposed a fast road monitoring model for UAVs, SAC-RSM, which achieves an inference speed of 38.3 FPS after quantization and acceleration using the Huawei Ascend CANNs. However, these methods need to consider inference efficiency further under more limited UAV computing resources with higher efficiency. Min et al.\cite{ref33} and Zhao et al.\cite{ref34} have proposed lightweight object detection networks with sub-bit level parameter sizes for real-time UAV applications, which also challenges the model's adaptability.

Researchers have recently developed various AI UAV systems based on computer vision, which have wide-ranging application value and reference significance levels. Li et al.\cite{ref35} developed a UAV-based object detection system that deploys Tiny-YOLO on mobile devices to detect targets captured by UAVs. Wu et al.\cite{ref36} achieved an FPS of 49.7 when detecting minor insulator defects on edge computing devices with the help of their proposed multiscale feature interaction-based transformer network (MFITN). The parameter counts of similar MFITN models\cite{ref37} lead to exponential increases in the computational costs incurred on edge devices with high-resolution images. EdgeAnt effectively addresses this issue, balancing accuracy and efficiency in edge inference tasks with minimal model parameters.

\begin{figure}[]
\centering
\includegraphics[width=3.5in, height=2.15in]{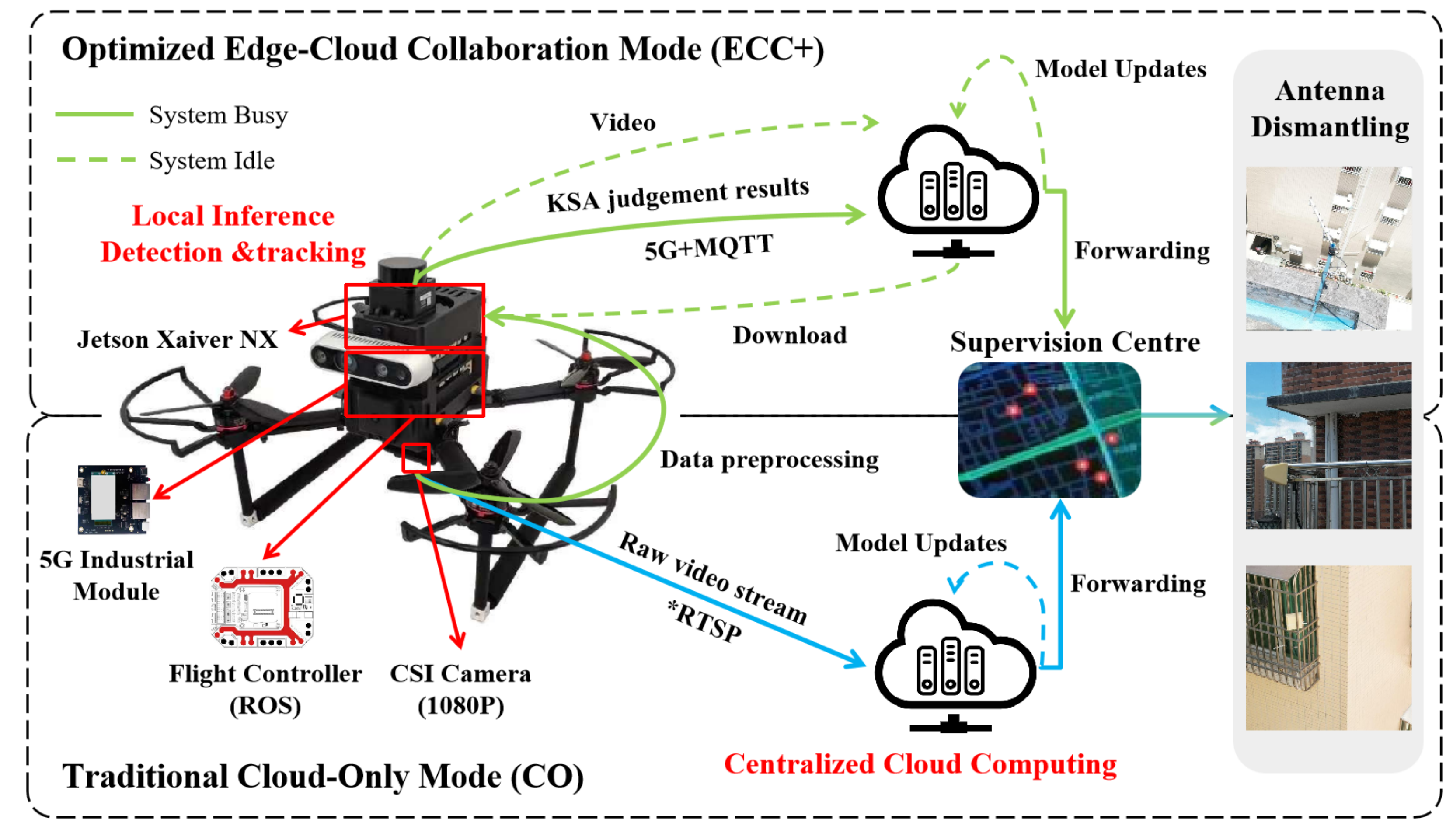}
\caption{Two different working paths for antenna interference source inspection systems: our ECC+-mode (green path) and the CO-mode (blue path).}
\label{fig3}
\end{figure}
    
\section{System platform and algorithm}
This section describes our AIoT system for UAV antenna interference source detection. The main contents are as follows. Section A introduces the system framework based on the ECC+ mode, including the KSA, and compares it with the mainstream co mode. Section B explores the system's scalability by modeling common scenarios involving multiple UAVs and large-scale joint inspections. Section C details EdgeAnt, a lightweight and efficient detector designed for edge computing scenarios, which is the core innovation of this paper. Finally, Section D briefly describes our improved AntSort tracker. 

\subsection{System Design and Comparison}
In the CO mode application, the edge device is a video acquisition tool that employs a GStreamer to stream the raw video via RTSP to the server. After the server receives the video stream, it feeds it into the recognition network to obtain the interference source localization results. It forwards the results to the monitoring center. Compared with the ultra-high bandwidth usage and delay caused by real-time video stream uploading in the CO mode, ECC+ achieves data uploads only through the lightweight IoT protocol MQTT. Fig. 3 shows the solutions yielded by our ECC+ mode and the CO mode in the interference source inspection task. Therefore, we define the E2EL in both computational modes as the sum of the data upload communication and inference time.

\renewcommand{\thealgorithm}{1}
    \begin{algorithm}
        \caption{\textcolor{black}{Keyframe Selection Algorithm in the ECC+ Mode}}
        \begin{algorithmic}[1]

            \Require A video sequence $\mathcal{F}=\left\{f_1, \ldots, f_N\right\}$; detector $Det$; tracker $Tra$; a pixel threshold $\tau$; a tracking threshold $\mu$
            \Ensure Inference result keyframes $\mathcal{K}$

            \State $Judge \gets \text{defaultdict}(\lambda: [0, 0, 0])$

            \For{$frame $ $ f_i $ $ in $ $\mathcal{F}$}  // Performing detection at the edge
                \State $\mathcal{D}_k \leftarrow Det(f_k);$
                \State $\mathcal{D}_{true} \leftarrow \emptyset;$ $\mathcal{T}_{k} \leftarrow \emptyset;$

                \For{$d $ $in $ $\mathcal{D}_k$} // First filtration step

                    \If {$d.w \leq \tau$ and $d.h \leq \tau$}
                    \State $\mathcal{D}_{true} \leftarrow \mathcal{D}_{true} \cup \{d\};$
                \EndIf
                \EndFor

                \State  $\mathcal{T}_{k} \leftarrow \emptyset;$
                \For{$d $ $in $ $\mathcal{D}_{true}$} // Performing tracking at the edge
                    \State $t \gets Tra(d);$

                    \If {$t.id $ $not$ $ in$ $ Exist_{id}$}
                        \State $\mathcal{T}_k \gets \mathcal{T}_k \cup \{t\};$
                    \EndIf
                \EndFor

                \For{$t $ $in $ $\mathcal{T}_k$} // Second filtration step
                    \State $Judge[t.id][0] \gets judge[t.id][0]+1;$
                    \If {$Judge[t.id][0] == 1$}
                        \State $Judge[t.id][1]=t.id;$
                    \EndIf
                    \State $Judge[t.id][2]=i;$

                    \If {$(Judge[t.id][0] >= \mu)$ and $(Judge[t.id][2] - Judge[t.id][1] <= (\mu-1))$}
                        \State $\mathcal{K} \gets \mathcal{K} \cup \{f_i\};$
                        \State $Exist_{id} \gets Exist_{id} \cup \{t.id\};$
                    \EndIf

                \EndFor

            \EndFor
            \State \Return $\mathcal{K}$
        \end{algorithmic}
    \end{algorithm}

\begin{figure*}[]
\centering
\includegraphics[width=5.2in]{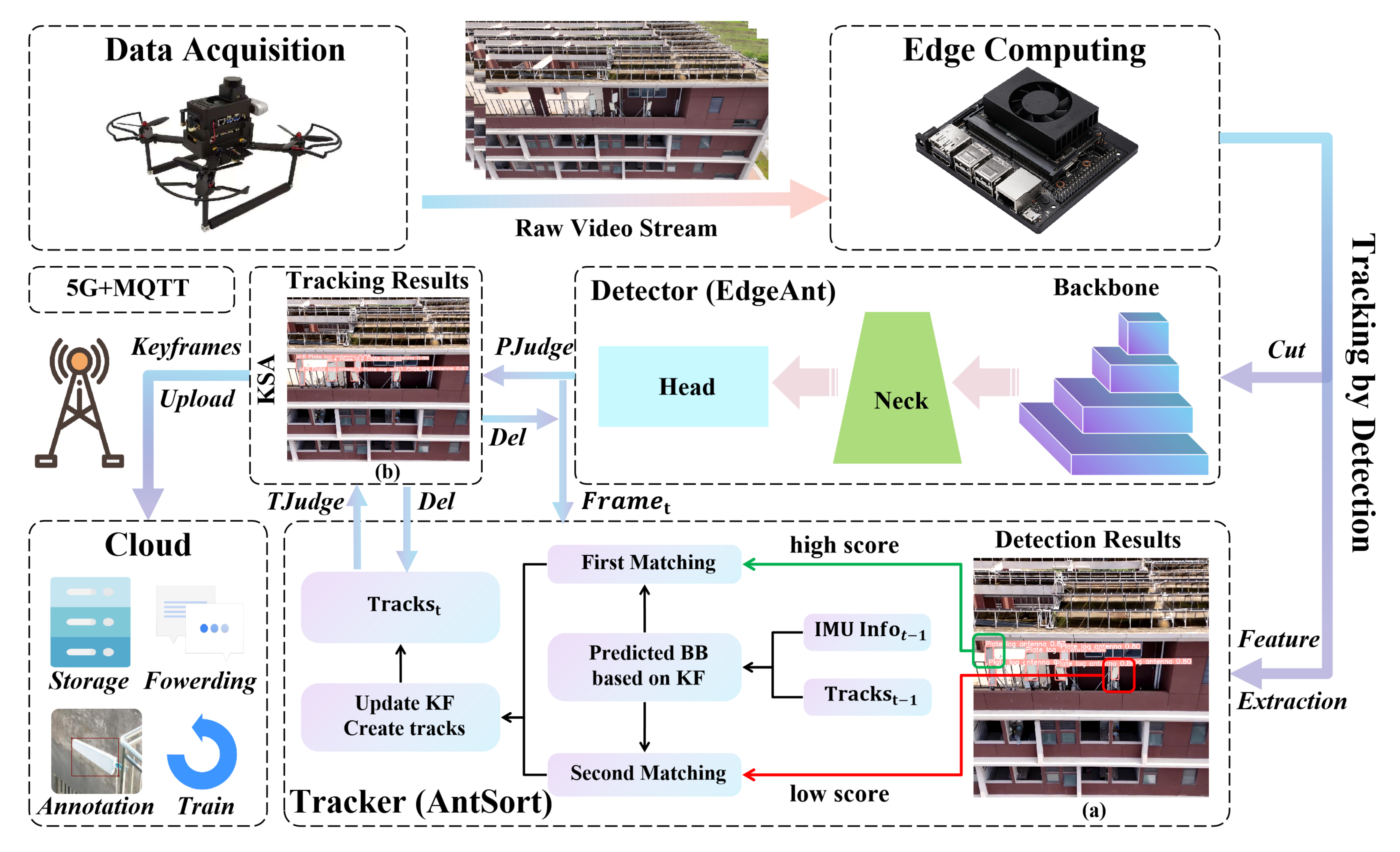}
\caption{Workflow diagram of the AIoT system based on antenna interference source inspection in the ECC+ mode. Notably, Fig. (a) shows the inference results of the detector, while Fig. (b) displays the tracking keyframes after the final KSA filtering. PJudge refers to pixel filtering of detection results, while TJudge refers to filtering of tracking result labels. The newly annotated dataset will retrain the detector on the cloud.}
\label{fig4}
\end{figure*}

\begin{equation}
\small
\mathrm{T}=\mathrm{T}^{\mathrm{C}}+\mathrm{T}^{\mathrm{I}}+\eta
\end{equation}
where $\mathrm{T}^C$ represents the communication time and $\mathrm{T}^I$ represents the inference time. The delay error $\eta$ includes the hardware delay of the camera sensor itself, the delay error caused by an unstable 5G signal during the communication process, and the data forwarding time. Notably, with assistance from the KSA in ECC+ mode, communication between the edge and the cloud involves only the instantaneous upload of localization results, with communication latency significantly lower than that of the CO mode.

Fig. 4 shows the complete workflow of our system. Our system consists of a cloud server with rich computing power, an edge computing device (Jetson Xavier NX) with relatively limited computing power, and an industrial mobile UAV platform. In the application, the real-time video stream captured by the camera is directly transmitted to the edge computing device for local inference via the deployed TBD paradigm-based localization algorithm. With the help of KSA, only key results are uploaded to the cloud server via mobile 5G modules. The raw videos are saved locally and uploaded during free time to further train and update the model.

Unlike most previous IoT applications in the ECC mode, we offload all detection and tracking tasks to the edge server, eliminating communication delays during cloud processing. The E2EL of the system depends solely on edge inference and result reporting times. To reduce the degree of data redundancy in video surveillance, we introduce the KSA to minimize edge-cloud communication, as shown in Algorithm 1. Considering the sizes of antenna targets in UAV-captured aerial images, we filter the results of the detector with a pixel threshold $\tau$. Objects larger than $\tau$ are not sent to the tracker. To ensure that complete interference source information is obtained, we set a tracking threshold $\mu$, where only targets that are tracked successfully for more than $\mu$ consecutive frames are deemed existing sources. These objects are uploaded once, and the local tracking records are deleted to reduce the burden imposed on the network. The processes for selecting $\tau$ and $\mu$ values are detailed in Section IV.

Our system aims to reduce the E2EL defined by (1). While conducting local processing with lightweight models on edge devices improves latency, this strategy often sacrifices accuracy compared with that of complex cloud models. To balance latency and inspection accuracy in real-time IoT applications, we propose EdgeAnt, a lightweight single-stage detector based on the TBD paradigm, and AntSort, an optimized tracker for real-time antenna interference source localization. We will cover this in more detail in Section C and D.

\subsection{Collaboration of Multiple UAVs}
Before mission commencement, intelligent industrial UAVs require the strategic planning of efficient flight paths to mitigate the endurance limitations imposed by high power consumption. By deploying multiple UAVs within the designated monitoring area, inspections of interference sources can be completed within a constrained timeframe, constituting a typical scan coverage problem\cite{ref38}, as illustrated in Fig. 5.

\begin{figure}[]
\centering
\includegraphics[width=3.4in, height=3.1in]{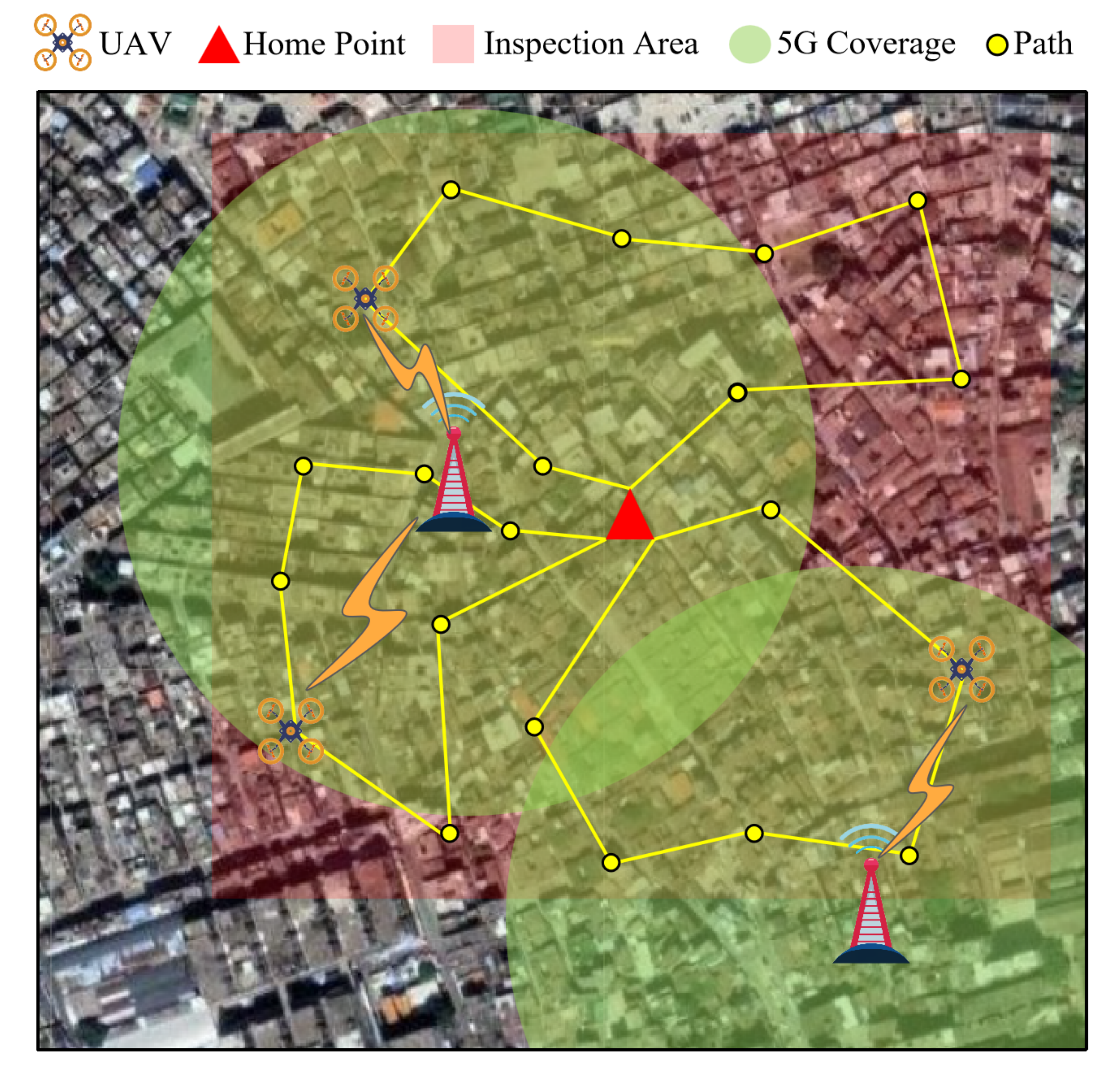}
\caption{A diagram of multiple UAVs conducting a coordinated inspection. The area is covered by 5G signals, with two base stations shown. UAVs return to the launch point after completing the inspection.}
\label{fig5}
\end{figure}

We formulate the inspection coverage problem for the designated area as a shortest-path maximum coverage problem, aiming to deploy UAVs that cover the most significant possible inspection area with minimal energy consumption. Additionally, the UAVs should remain as close as possible to the base station during flight to minimize E2EL. Accordingly, the objective function can be defined as follows

\begin{equation}
\small
\text J=\sum_{j=1}^n\left(\text { L}_j+\alpha_1\sum_{k=1}^m \frac{1}{\text { FSPL}\left(P_k, x_j, y_j\right)}+\alpha_2S_j+\alpha_3\widetilde{A}\right)
\end{equation}
where n represents the number of UAVs, m denotes the number of base stations covering the inspection area. $\alpha_1$, $\alpha_2$, and $\alpha_3$ are weighting coefficients. $L_j$ refers to the path length of the j UAV. Free Space Path Loss (FSPL) indicates path loss directly affects communication quality, resulting in reduced data transmission rates and increased signal retransmissions. Each UAV selects the nearest base station for communication handover at each node along its path. $S_j$ represents the number of base station handovers for the j UAV, and $A_u$ represents the uncovered area.

We divide the inspection area of size $L \times L$ into $N$ smaller grids, where $A_{\text {g }}=\Delta^2$ represents the area of each grid cell. The coverage radius of each UAV is denoted as $R$. By grid partitioning the inspection area; the uncovered area can be determined as

\begin{equation}
\small
\widetilde{A} =\sum_{k=1}^{N_{\text {g }}}\left[\left(\min _i\left(\sqrt{\left(x_i-x_k\right)^2+\left(y_i-y_k\right)^2}\right)>R\right)\right] \times A_{\text {g}}
\end{equation}
where $\left(x_k, y_k\right)$ represents the midpoint of each grid. Each UAV has edge computing devices and mobile modules, enabling communication with ground base stations via air-to-ground (A2G) wireless links. Compared to air-to-air (A2A) channels, A2G channels are more prone to significant shadowing and small-scale fading, leading to increased packet loss and retransmissions, adversely impacting communication latency and quality. Given the limited distance, signal propagation delay is negligible. The FSPL model is used to describe signal attenuation.

\begin{equation}
\small
\operatorname{FSPL}=20 \log _{10}(d)+20 \log _{10}(f)+20 \log _{10}\left(\frac{4 \pi}{c}\right)
\end{equation}
where FSPL represents the path loss, $d$ is the signal propagation distance, $f$ is the signal carrier frequency, and $c$ is the speed of light.

Particle Swarm Optimization (PSO), inspired by bird flocks' foraging behavior, allows each particle to rely on its own experience and the information from other particles, effectively avoiding the problem of getting trapped in local optima. In this context, each particle represents the path of a UAV.

\begin{equation}
\small
\operatorname{P}_j=\left(\left(x_{j 1}, y_{j 1}, \theta_{j 1}\right), \ldots,\left(x_{j n}, y_{j n}, \theta_{j n}\right)\right)
\end{equation}
where $\left(x_\text{ji}, y_\text{ji}\right)$ represents the coordinates of the UAV, and $\theta$ denotes the flight direction. Initially, all UAVs start from the center of the inspection area $(L / 2, L / 2)$, with flight directions uniformly distributed. After completing the inspection, they return to the starting point. The velocity update equation is

\begin{equation}
\small
v_{j i}(t+1)=\omega v_{j i}(t)+c_1\left(p_{j i}^{\text {b }}-x_{j i}(t)\right)+c_2\left(p_{j i}^{\text {g }}-x_{j i}(t)\right)
\end{equation}
where $p_b$ represents the particle's personal best position and direction, while $p_g$ denotes the global best position and direction, $c_1$ and $c_1$ are random factors and learning factors, respectively, which control the step size towards the individual and globally optimal solutions. $w$ is the inertia weight. The position update formula is given by

\begin{equation}
\small
\left\{\begin{array}{l}
x_{j i}(t+1)=x_{j i}(t)+v_{x, j i}(t+1) \\ [0.4em]
y_{j i}(t+1)=y_{j i}(t)+v_{y, j i}(t+1) \\ [0.4em]
\theta_{j i}(t+1)=\theta_{j i}(t)+\Delta \theta_{j i}(t+1)
\end{array}\right.
\end{equation}

Finally, the updated flight direction is used to calculate the next movement of each particle.

\begin{equation}
\small
\left\{\begin{array}{l}
x_{j i}(t+1)=x_{j i}(t)+v_{j i}(t+1) \cdot \cos \left(\theta_{j i}(t+1)\right) \\ [0.4em]
y_{j i}(t+1)=y_{j i}(t)+v_{j i}(t+1) \cdot \sin \left(\theta_{j i}(t+1)\right)
\end{array}\right.
\end{equation}

When scaling up for large-scale inspections, UAVs with independent communication can form a mesh network to improve A2G communication. Edge computing can assist with complex path planning. However, using the same frequency band for hundreds of UAVs can cause interference. Dynamic spectrum allocation and interference-aware protocols can optimize communication by maximizing the Signal-to-Interference-plus-Noise Ratio (SINR) to address this.

\begin{equation}
\small
\operatorname{SINR}=\frac{P_s G_s\left(\frac{\lambda}{4 \pi d_s}\right)^2}{I+N_0},I=\sum_{i=1}^{N_{\text {intf }}} P_i G_i\left(\frac{\lambda}{4 \pi d_i}\right)^2
\end{equation}

\begin{equation}
\small
J^{\prime}=J+\alpha_4 \sum_{j=1}^n\left(\frac{1}{\operatorname{SINR}_j}\right)
\end{equation}
where $I$ is the interference power, $P_\textbf{i(s)}$ is the transmission power of the i(s)-th interfering (servicing) UAV, $G_\text{i(s)}$ is the antenna gain, $\lambda$ is the wavelength, and $d_\textbf{i(s)}$ is the distance between the UAVs. $N_0$ is the noise power spectral density. $\alpha_4$ is the weight coefficient, and $\text{SINR}_j$ is the SINR of the j-th UAV.

In addition, to coordinate more complex path planning, a collision avoidance mechanism is introduced into the PSO.

\begin{equation}
\small
V_{\mathrm{ca}}=\sum_{i=1}^n \sum_{j=i+1}^n \frac{C}{\left|\mathbf{p}_i-\mathbf{p}_j\right|^q} \end{equation}

\begin{equation}
\small
J^{\prime\prime}=J^{\prime}+\alpha_5 V_{\text {ca }}
\end{equation}
where C is a constant, $\mathbf{p}_i$ and $\mathbf{p}_j$ are the positions of the i-th and j-th UAVs, and q is the exponent that determines the strength of the repulsive force between UAVs.

\subsection{EdgeAnt}
The detector is essential for localizing and recognizing antenna objects in video frames; its accuracy determines the effectiveness of subsequent tracking and KSA determinations. We face the challenge of achieving high-speed operations while maintaining strong detection accuracy on resource-limited edge devices, particularly for small targets. To address this issue, we propose the EdgeAnt detector, which balances the number of model parameters, detection speed, and accuracy. The specific architecture is shown in Fig. 6.

\begin{figure*}[]
\centering
\includegraphics[width=5.1in, height=3.0in]{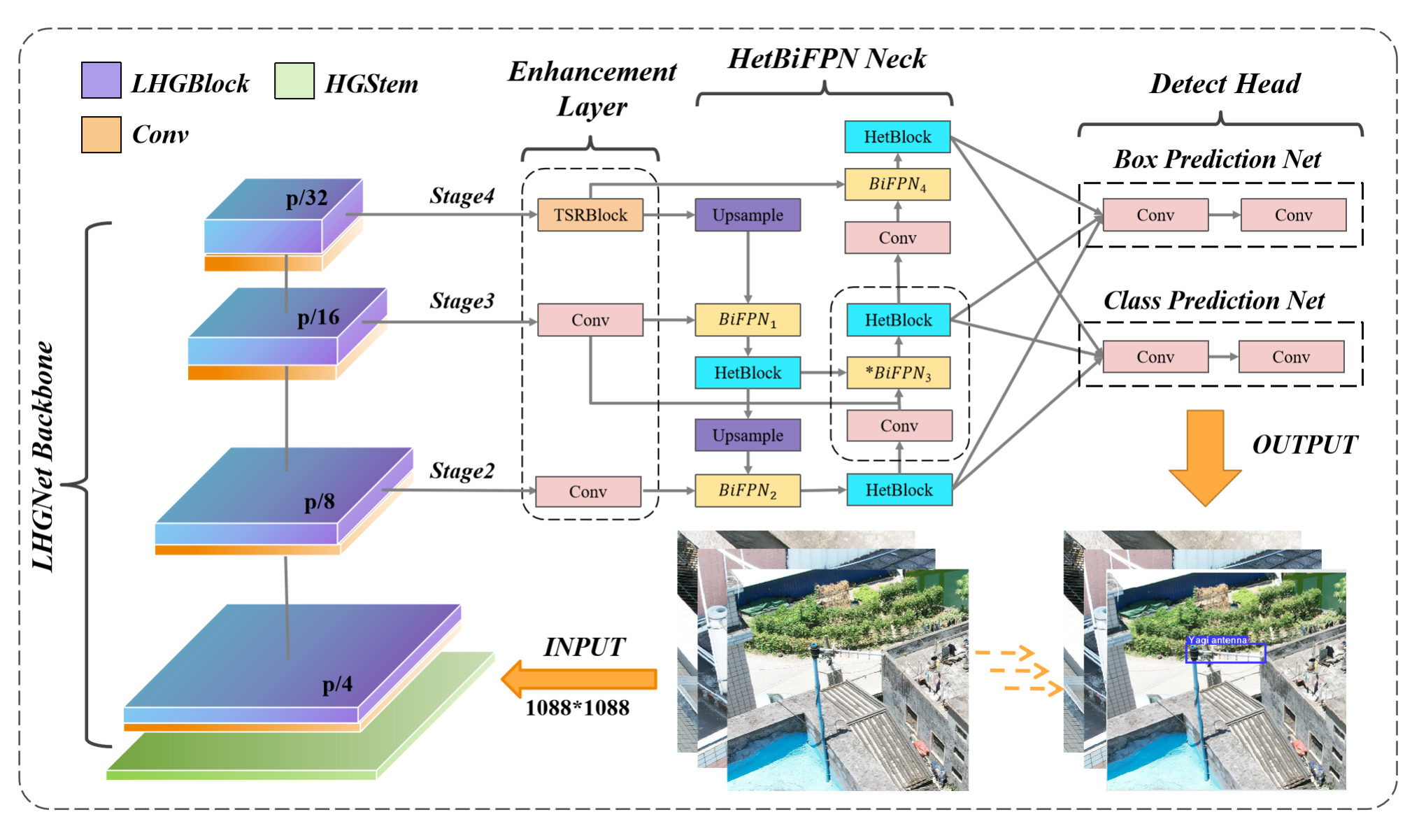}
\caption{The comprehensive architectural diagram of the EdgeAnt detector illustrates its composition: the feature extraction backbone lightweight hierarchical geometric network (LHGNet), the feature fusion neck the heterogeneous bidirectional feature pyramid network (HetBiFPN), the small object enhancement layer (EL), and the classification and regression detection heads. The EL acts as a bridge between the backbone and neck networks, effectively minimizing the loss of features related to small objects.}
\label{fig6}
\end{figure*}

EdgeAnt, a new baseline single-stage detector, features a three-in-one architecture integrating a feature extraction and enhancement network. Table I provides a detailed breakdown of the composition of EdgeAnt and YOLOv10-n at each level. It can be observed that EdgeAnt directly downsamples the input image by a factor of 4 to reduce the computational load of the model. Consequently, more channels are utilized to compensate for the loss of image features. We perform a lightweight reconstruction process on HGNetv2\cite{ref39} to maintain strong performance for complex images while minimizing the number of redundant computations, resulting in a new backbone: LHGNet. Inspired by PANet\cite{ref40} and BiFPN\cite{ref41}, we implement a dual-path and tri-path mixed feature fusion mechanism in the neck, incorporating a core component called HetBlock to form the HetBiFPN neck architecture. Furthermore, we have eliminated using traditional spatial pooling operations to capture multi-scale features. EL is added at the junction between the backbone and neck to leverage the proposed TSRBlock and enhance the ability to capture information about small targets in real-time detection tasks.

\begin{table}[]
\centering
\caption{COMPARISON OF ARCHITECTURE SPECIFICATIONS BETWEEN EDGEANT AND YOLOV10-N}
\label{tab:my-table}
\renewcommand{\arraystretch}{1.2}
\resizebox{\columnwidth}{!}{
\begin{tabular}{ccc|ccc}
\hline
\multicolumn{3}{c|}{EdgeAnt}                                & \multicolumn{3}{c}{YOLOv10-n}                                   \\ \hline
F\&L        & Input                      & Block    & F\&L      & Input                           & Block      \\ \hline
\multicolumn{3}{c|}{Backbone}                               & \multicolumn{3}{c}{Backbone}                                    \\ \hline
$0$ {[}$0${]}               & 1088$^2 \times 3$  & HGStem   & $0$ {[}$0${]}            & 1088$^2 \times 3$       & Conv       \\
$1$ {[}$-1${]}              & 272$^2 \times 64$  & Conv     & $1$ {[}$-1${]}           & 544$^2 \times 16$       & Conv       \\
$2$ {[}$-1${]}              & 136$^2 \times 32$  & LHGBlock & $2$ {[}$-1${]}           & 272$^2 \times 32$       & C2f        \\
$3$ {[}$-1${]}              & 136$^2 \times 128$ & Conv     & $3$ {[}$-1${]}           & 272$^2 \times 32$       & Conv       \\
$4$ {[}$-1${]}               & 68$^2 \times 64$   & LHGBlock & $4$ {[}$-1${]}           & 136$^2 \times 64$       & C2f        \\
$5$ {[}$-1${]}              & 68$^2 \times 256$  & Conv     & $5$ {[}$-1${]}           & 136$^2 \times 64$       & SCDown     \\
$6$ {[}$-1${]}              & 34$^2 \times 128$  & LHGBlock & $6$ {[}$-1${]}           & 68$^2 \times 128$       & C2f        \\
$7$ {[}$-1${]}              & 34$^2 \times 512$  & Conv     & $7$ {[}$-1${]}           & 68$^2 \times 128$       & SCDown     \\
$8$ {[}$-1${]}              & 17$^2 \times 128$  & LHGBlock & $8$ {[}$-1${]}           & 34$^2 \times 256$       & C2f        \\ \cline{1-3}
\multicolumn{3}{c|}{Enhancement Layer}                      & $9$ {[}$-1${]}           & 34$^2 \times 256$       & SPPF       \\ \cline{1-3}
$9$ {[}$4${]}               & 68$^2 \times 256$  & Conv     & $10$ {[}$-1${]}          & 34$^2 \times 256$       & PSA        \\ \cline{4-6}
$10$ {[}$6${]}              & 34$^2 \times 512$  & Conv     & \multicolumn{3}{c}{Neck}                                        \\ \cline{4-6}
$11$ {[}$8${]}              & 17$^2 \times 512$  & TSRBlock & $11$ {[}$-1,6${]}  & 68$^2 \times 384$ & *Concat    \\ \cline{1-3}
\multicolumn{3}{c|}{Neck}                                   & $12$ {[}$-1${]}          & 68$^2 \times 384$       & C2f        \\ \cline{1-3}
$12$ {[}$-1,10${]} & 34$^2 \times 64$ & *BiFPN   & $13$ {[}$-1,4${]}                 & 136$^2 \times 192$ & *Concat                 \\
$-1$ {[}$13${]}             & 34$^2 \times 64$   & HetBlock & $14$ {[}$-1${]}          & 136$^2 \times 192$      & C2f        \\
$14$ {[}$-1,9${]}     & 68$^2 \times 64$   & *BiFPN   & $15$ {[}$-1${]}          & 136$^2 \times 64$       & Conv \\
$15$ {[}$-1${]}             & 68$^2 \times 64$   & HetBlock & $16$ {[}$-1,12${]} & 68$^2 \times 192$  & Concat     \\
$16$ {[}$-1${]}             & 68$^2 \times 64$   & Conv     & $17$ {[}$-1${]}          & 68$^2 \times 192$       & C2f        \\
$17$ {[}$-1,10,13${]} & 34$^2 \times 64$   & BiFPN    & $18$ {[}$-1${]}          & 68$^2 \times 128$       & SCDown     \\
$18$ {[}$-1${]}             & 34$^2 \times 64$   & HetBlock &  $19$ {[}$-1,10${]} & 34$^2 \times 384$ & Concat     \\
$19$ {[}$-1${]}             & 34$^2 \times 128$  & Conv     & $20$ {[}$-1${]}          & 34$^2 \times 384$       & C2fCIB     \\ \cline{4-6}
$20$ {[}$-1,11${]}    & 17$^2 \times 64$   & BiFPN    & \multicolumn{3}{c}{Head}                                        \\ \cline{4-6}
$21$ {[}$-1${]}          & 17$^2 \times 64$ & HetBlock & \multirow{3}{*}{$21$ {[}$14,17,20${]}} & \multirow{3}{*}{-}              & \multirow{3}{*}{Detect} \\ \cline{1-3}
\multicolumn{3}{c|}{Head}                                   &                  &                                 &            \\ \cline{1-3}
$22$ {[}$15,18,21${]}      & -                          & Detect   &                  &                                 &            \\ \hline
\end{tabular}%
}
\end{table}

\subsubsection{LHGNet Backbone}
Detecting antenna interference sources is challenging because they appear very small in images and can be obscured by factors like background noise and lighting. The HGNetv2\cite{ref39} backbone uses a method that extracts features at multiple levels, enhancing the network's ability to learn complex patterns of different sizes. However, its high computational complexity makes it unsuitable for real-time detection tasks in mobile UAV scenarios. To address this issue, we propose a lightweight backbone network called LHGNet, designed to minimize redundant computations while retaining effective feature extraction capabilities.

The process of reducing the weight of the original backbone can be divided into two steps. The first step involves reducing the weight of the core component (HGBlock) of the backbone, and the second step involves adjusting the depth and structure of the entire backbone. LHGBlock is designed for hierarchical data processing. To simplify the model parameters, an additional 1x1 convolution is used for dimensionality expansion during the output step in HGBlock. We remove this additional convolution to simplify the calculation process and speed up the forward propagation step. By doing so, the feature maps in the output channels also possess more local features accordingly. When stacking bottlenecks in HGBlock, HGNetv2\cite{ref39} selectively uses LightConv. LightConv replaces the second of two standard convolutions in a conventional stack with a depthwise convolution (DWConv\cite{ref42}). Utilize DWConv\cite{ref42} at critical points in the core components is unwise, as its inherent inability to fully capture cross-channel feature relationships will lead to feature losses. Therefore, we continue replacing all standard convolutions in HGBlock with GhostConv\cite{ref43}, as shown in (13). Although this may lead to a slight parameter increase, it is worthwhile.

\begin{equation}
\small
\mathrm{Y}=\operatorname{cat}\left(\mathrm{x}, \operatorname{cat}\left(\sum_{i=1}^n \mathrm{y}_{\mathrm{i}}\right)\right), \mathrm{y}_{\mathrm{i}}=\mathrm{f}_{\mathrm{GS}}^{\mathrm{i}}(\mathrm{x})
\end{equation}
where $\mathrm{f}_{\mathrm{GS}}^{\mathrm{i}}$ represents the input feature map obtained after $\mathrm{i}$ iterations of the GhostConv operation.

\begin{figure}[]
\centering
\includegraphics[width=3.5in, height=2.7in]{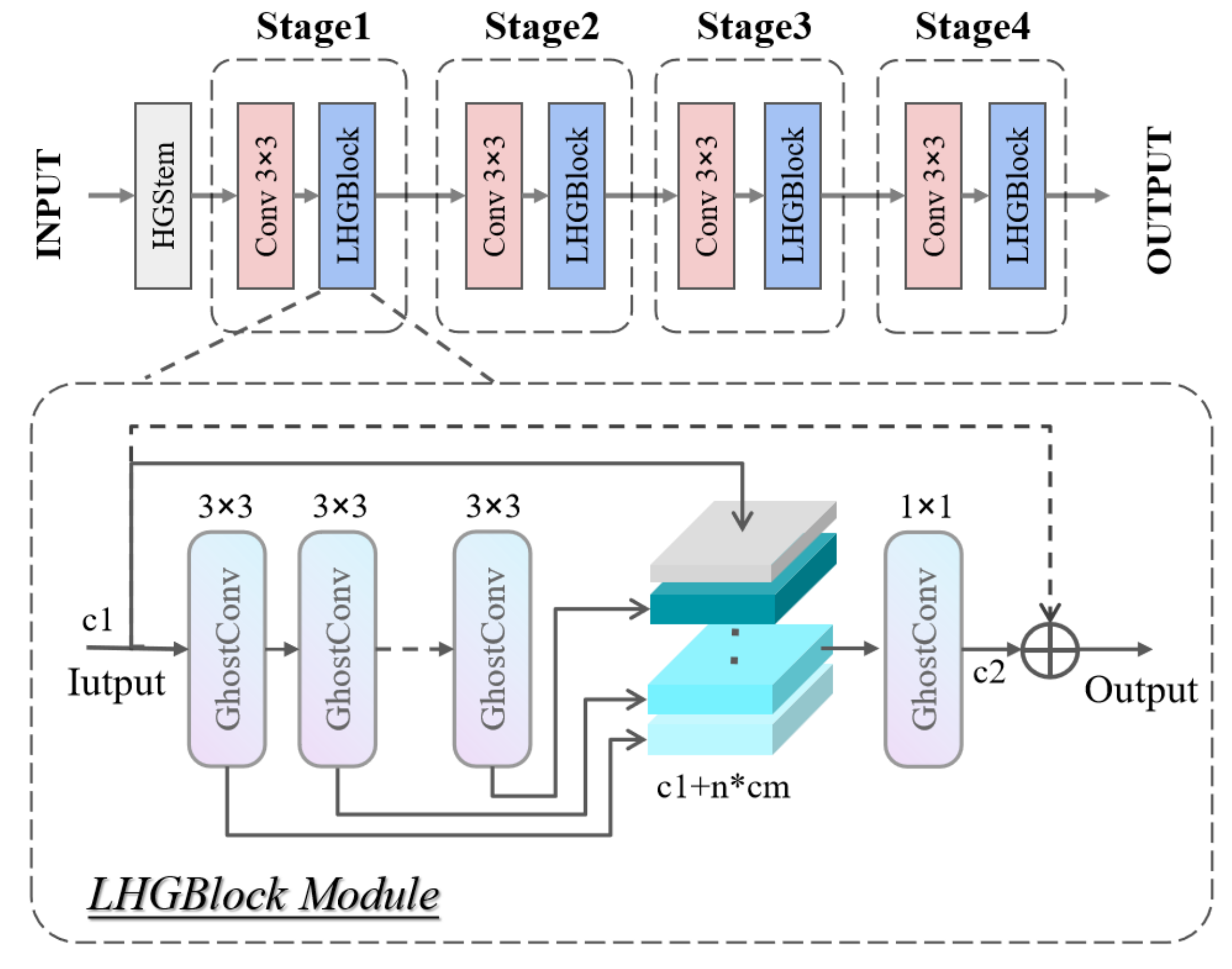}
\caption{Structure of LHGNet. Redundant computations are minimized in the backbone through pruning and modular reconstruction methods.}
\label{fig7}
\end{figure}

Fig. 7 shows the LHGNet backbone that we finally construct. The most crucial adjustment is that after the input feature map goes through HGStem initialization processing, it is directly passed into a standard convolution. This operation reduces the number of floating-point operations (FLOPs) by nearly four times with almost no change in the number of model parameters. In other words, we conduct two preliminary feature extraction operations on the input features, mapping the channels to a number of channels that is more suitable for the subsequent core components. The feature map that has just entered the model often contains the richest target details, and owing to the dual extraction steps performed by the employed 3x3 convolutions, the original details of the image are preserved to the greatest extent possible. The feature accuracy loss and the decrease in the computational cost form an effective tradeoff. We prune each subsequent stage, retaining only the 3x3 convolution for downsampling and the core feature extraction component (LHGBlock).

\subsubsection{HetBiFPN Neck}
The neck structure is designed to extract and combine features obtained from the backbone. Traditional top-down FPN\cite{ref44} is limited because it only allows information to flow in one direction. PANet\cite{ref40} adds a bottom-up path to solve this issue, but its simple dual-path feature fusion increases the model's computational cost. BiFPN\cite{ref41} lets information flow and merge in both top-down and bottom-up directions through a weighted fusion mechanism, allowing the network to use information from different scales effectively. However, adding too many fusion links at lower levels in small object detection tasks can introduce ineffective features. Inspired by this idea, we stick to the dual-path fusion mechanism of PANet\cite{ref40} to capture features from relatively low stages of the network. We also incorporate a weighting mechanism to balance contributions of different sizes during fusion. Only in $BiFPN_3$ do we allow the input of additional mid-stage network features to detect small objects better.

C2f, as the key to the success of YOLOv8, guarantees network performance by implementing the ResNet idea of splitting channels and connecting them with multilevel residuals. Moreover, we aim to further lighten this design. Compared with model pruning, heterogeneous kernel-based convolution (HetConv\cite{ref45}) reduces the numbers of computations and parameters while maintaining high representation efficiency, and it has been experimentally proven to be effective. Therefore, we propose a lightweight feature extraction module, HetBlock, for neck construction purposes. Fig. 5 shows the final complete neck architecture (HetBiFPN), with the details of HetBlock illustrated in Fig. 8.

\begin{figure}[]
\centering
\includegraphics[width=3.5in]{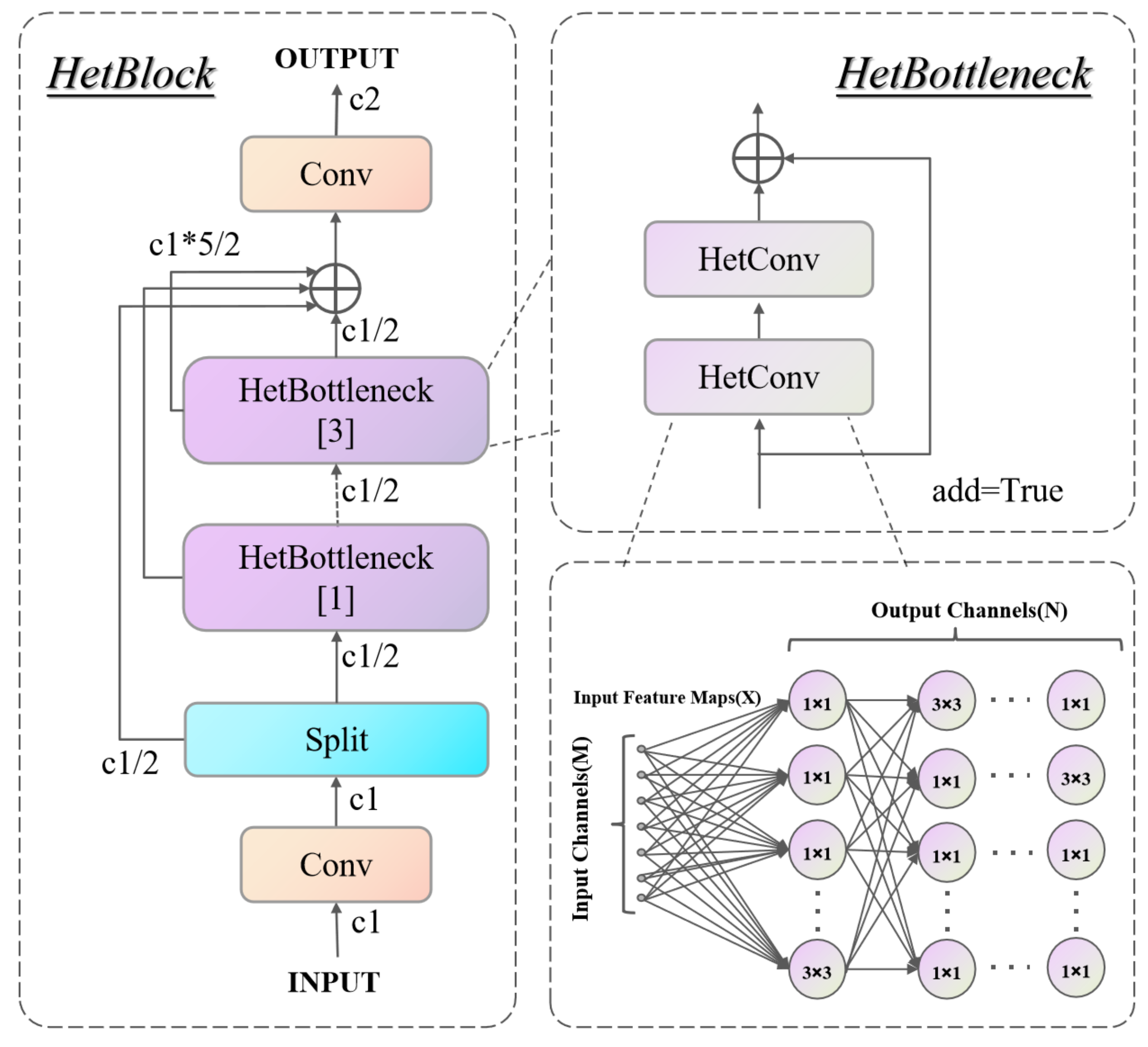}
\caption{Structure of HetBlock. The module makes full use of $1\times1$ and $3\times3$ convolutional kernels.}
\label{fig8}
\end{figure}

Specifically, we fix the number of bottlenecks in C2f to 3, where the bottlenecks are constructed via HetConv. We implement the HetConv concept to replace three-quarters of the convolution kernels with $1\times1$ convolution kernels in a standard $3\times3$ convolution operation, which performs convolution operations asymmetrically with the original $1\times1$ convolution kernels. This reduces the computational cost and number of parameters while maintaining the representation efficiency of the CNN. Assuming that the input feature map in a single HetConv operation is denoted as $\mathrm{W}_{\mathrm{i}} \times \mathrm{H}_{\mathrm{i}} \times \mathrm{C}_{\mathrm{i}}$, where W and H represent the width and height of the input feature map, respectively, and M represents the number of input channels, the output is also denoted as $\mathrm{W}_{\mathrm{o}} \times \mathrm{H}_{\mathrm{o}} \times \mathrm{C}_{\mathrm{o}}$. The standard convolution operation uses A filters with sizes of $3\times3\times C_i$ to produce the output feature map, where 3 represents the convolution kernel size. Therefore, the total computational cost of performing this convolution operation once can be expressed as

\begin{equation}
\small
\mathrm{F}_{\text {Conv }}^{\text {Cost }}=\mathrm{W} \times \mathrm{H} \times \mathrm{C}_{\mathrm{i}} \times \mathrm{C}_{\mathrm{o}} \times 3 \times 3
\end{equation}

In HetConv, half of the convolution kernels are replaced with $1\times1$ kernels, and they are alternately arranged with the remaining half of the $3\times3$ convolution kernels to form a filter. The computational cost of these $3\times3$ convolution kernels is

\begin{equation}
\small
\mathrm{F}_{\text {HetConv }}^{\text {Cost1 }}=\frac{\mathrm{W} \times \mathrm{H} \times \mathrm{C}_{\mathrm{i}} \times \mathrm{C}_0 \times 3 \times 3}{4}
\end{equation}

The computational cost of the remaining half of the $1\times1$ convolution kernels is

\begin{equation}
\small
\mathrm{F}_{\text {HetConv }}^{\text {Cost2 }}=\frac{\mathrm{W} \times \mathrm{H} \times \mathrm{C}_{\mathrm{i}} \times \mathrm{C}_0}{\frac{4}{3}}
\end{equation}

Therefore, the total cost of a single HetConv operation is

\begin{equation}
\small
\mathrm{F}_{\text {HetConv }}^{\text {Cost }}=\mathrm{F}_{\text {HetConv }}^{\text {Cost1 }}+\mathrm{F}_{\text {HetConv }}^{\text {Cost2 }}
\end{equation}

Compared with that of a standard bottleneck, the total computational cost reduction attained by a bottleneck constructed with HetConv can be represented as R.

\begin{equation}
\small
\mathrm{R}_{\text {Bottleneck }}^{\mathrm{n}=1}=12 \times \mathrm{W} \times \mathrm{H} \times \mathrm{C}_{\mathrm{i}} \times \mathrm{C}_{\mathrm{o}}
\end{equation}

The use of HetBlock minimizes the parameter count and computational complexity in the bottleneck to the greatest extent possible. However, HetConv retains one quarter of the alternating $3\times3$ convolution kernels, ensuring that the filters capture spatial correlations in specific channels. The receptive field and feature extraction performance of the module remain unchanged. Our calculation represents only the gain achieved by replacing a bottleneck.

\subsubsection{Enhancement Layer}
\begin{figure}[]
\centering
\includegraphics[width=3.5in,height=1.6in]{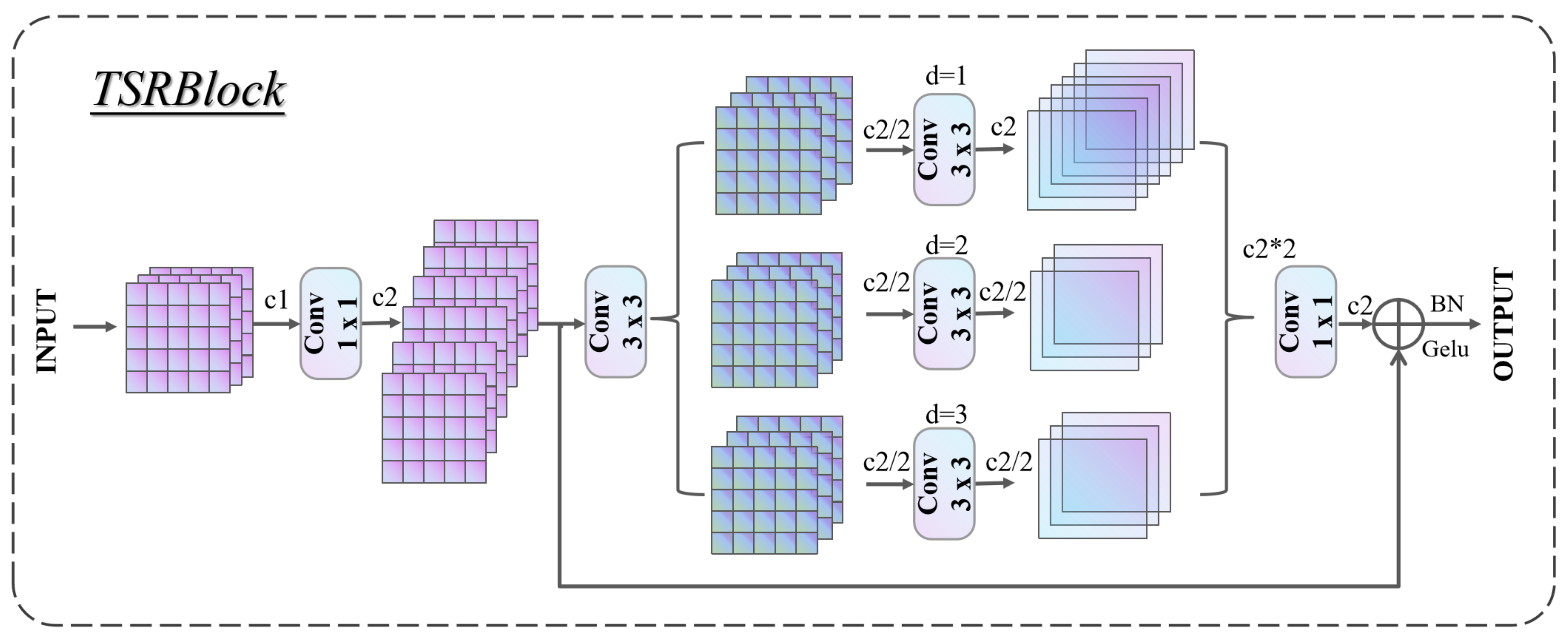}
\caption{Structure of the TSRBlock. Residual dilation strengthens the focus of the model on small targets.}
\label{fig9}
\end{figure}

Lightweight detectors that meet practical application requirements face challenges in small object detection tasks. This is partly because these models often use significant downsampling rates, especially in the early stages of feature extraction, making it hard for them to learn the features of small targets. Also, adding spatial pyramid pooling (SPP)\cite{ref46,ref47,ref48} between the backbone and neck has become essential for addressing image distortion and reducing computational costs. However, pooling operations can lead to some feature loss. Therefore, we remove and replace the SPP operations with an EL layer. This layer performs extra feature extraction on the shallower feature maps to help the network focus more on small objects. Large kernel convolutions\cite{ref49} can increase the receptive field but add extra parameters. Inspired by \cite{ref50}, we propose TSRBlock, which uses a two-step residual feature extraction method to effectively improve the network's ability to capture multiscale information in real-time object detection scenarios, as shown in Fig. 9.

TSRBlock precisely captures multiscale contextual information using a two-step approach to obtain detailed receptive fields, then fuses the feature maps from multiple scales. In the first step, DWR\cite{ref50} directly generates concise feature maps using a $3\times3$ convolution, followed by batch normalization (BN) and a rectified linear unit (ReLU). However, using a $3\times3$ convolution for channel expansion is highly complex, and we believe that normalization and activation aimed at speeding up model convergence may lead to some feature loss. So, we first perform a $1\times1$ convolution, allowing the $3\times3$ convolution to extract concise feature maps smoothly. We then move the normalization and activation operations to the end of the second step. In the second step, we use DWR operations, applying convolutions with different dilation rates to filter regional features. Each branch has a unique receptive field, forming a comprehensive feature representation. Given the small-target characteristics of antenna interference sources, we replace the dilation rate of 5 used in the convolution with a dilation rate of 2 to minimize redundancy in the receptive field. Finally, we compress the channels using a $1\times1$ convolution.

As shown in Fig. 5, we introduce TSRBlock only at the end of the EL to extract the output of the fourth stage of the backbone. This captures the feature values of inputs at different stages of the backbone, where the last layer is often the most valuable. By introducing the two-step residual method and fine-grained receptive field filter, TSRBlock optimizes the effectiveness of the multi-rate depthwise dilated convolution, constructing more robust and comprehensive feature representations. It provides a more accurate and robust foundation for detecting small interference sources while minimizing its impact on the detector's real-time inference speed.

\vspace{-5pt}
\subsection{AntSort}
The tracker relies on information from historical frames to generate target trajectories and associate detection boxes for tracking the target, ensuring continuity within the TBD paradigm. To satisfy the requirements of practical applications, we select BotSort\cite{ref51} as the baseline tracker. We adjust and optimize this tracker to obtain AntSort, which is a tracker designed explicitly for antenna target tracking. The information from the inertial measurement unit (IMU) effectively compensates for the target's state, enabling more accurate inspections. Specifically, AntSort directly utilizes the linear acceleration measurements acquired from the IMU of the employed UAV for motion camera compensation, updates and predicts the bounding box position of the target at the next moment, and omits the original random sample consensus (RANSAC) calculation to speed up the computation process. As shown in (19), the IMU data $\tilde{\mathrm{U}}_{\mathrm{k}-1}^{\mathrm{k}}$ are used as the input control variable.

\begin{equation}
\small
\tilde{\mathrm{U}}_{\mathrm{k}-1}^{\mathrm{k}}=\left[\mathrm{a}_{\mathrm{k} \mid \mathrm{k}-1}^{\mathrm{imux}}, \mathrm{a}_{\mathrm{k} \mid \mathrm{k}-1}^{\mathrm{imuy}}, \mathrm{a}_{\mathrm{k} \mid \mathrm{k}-1}^{\mathrm{imuz}}\right]
\end{equation}

\begin{equation}
\small
\hat{\mathrm{x}}_{\mathbf{k} \mid \mathbf{k}-1}^{\prime}=\tilde{\mathrm{F}}_{\mathrm{k}-1}^{\mathrm{k}} \hat{\mathrm{x}}_{\mathbf{k} \mid \mathrm{k}-1}+\tilde{\mathrm{U}}_{\mathrm{k}-1}^{\mathrm{k}}
\end{equation}
where $a_{k \mid k-1}^{imux}$, $a_{k \mid k-1}^{imuy}$, and $a_{k \mid k-1}^{imuz}$ represent the linear acceleration measurements provided by the IMU in different directions, and the compensated state is used for updating and prediction.

\begin{equation}
\small
\hat{\mathrm{x}}_{\mathrm{k} \mid \mathrm{k}}=\hat{\mathrm{x}}_{\mathrm{k} \mid \mathrm{k}-1}^{\prime}+\mathrm{K}_{\mathrm{k}}\left(\mathrm{z}_{\mathrm{k}}-\mathrm{H}_{\mathrm{k}} \hat{\mathrm{x}}_{\mathrm{k} \mid \mathrm{k}-1}^{\prime}\right)
\end{equation}

AntSort's specific workflow is shown in Algorithm 2, where Camera Motion Compensation (CMC) registers inter-frame targets. After tracking for each frame, KSA is executed to upload interference source targets that meet the criteria to the cloud and no longer match them in the future.

\renewcommand{\thealgorithm}{2}
\begin{algorithm}
\caption{Pseudo-code of AntSort}
\textbf{Input:} A video sequence $V$; detect results $D_k$; detection score threshold $\varepsilon$\\; feature extractor $Ext$; keyframe selection algorithm $KSA$

\textbf{Output:} Tracks $\mathcal{T}$ of the video
\begin{algorithmic}[1]
\State Initialize $\mathcal{T} \gets \emptyset$
\For{frame $f_k$ in $V$}
    \State Divide $D_{\text{high}}$ and $D_{\text{low}}$ by $\varepsilon$

    \State Predict new locations of tracks using Eq. 11
    \For{$t$ in $\mathcal{T}$}
        \State $t \gets$ $CMC($KalmanFilter($t$))
    \EndFor
    \State // Only extract high-confidence target features
    \State First associate $\mathcal{T}$ and $Ext$($D_{\text{high}}$) using IoU
    \State Second associate $\mathcal{T}$ and $D_{\text{low}}$ $\cup$ unmatched $D_{\text{high}}$ using IoU
    \State Delete unmatched and filter tracks
    \State $\text{KSA}(\mathcal{T}_\text{matched})$

    \State Initialize new tracks
\EndFor
\State \textbf{Return} $\mathcal{T}$
\end{algorithmic}
\end{algorithm}

When a UAV performs interference source detection tasks, it typically flies faster to improve its inspection efficiency. At this time, the image capture frequency is high, and AntSort utilizes the IMU of the UAV to compensate for the motion of the rigid camera. This effectively reduces the target position changes caused by camera movement, thereby enhancing the tracking accuracy and robustness of the system.

\section{Experiments}
This section outlines the preparations made for the experiments. The proposed system was evaluated by assessing the effectiveness of the detector through ablation experiments conducted on the backbone, neck, and EL, including the lightweight backbone design process. In our comparative experiments, we deployed SOTA detectors, both recent and classic, on edge computing devices (Jetson Xavier NX and Raspberry Pi 4B) for evaluation. We selected KSA parameters through experiments and compared system performance across different AI application modes and localization algorithms. Finally, we simulated the scalability of multiple UAVs cooperative inspection.

\subsection{Experimental Preparation}
\textbf{Datasets}
\subsubsection{Antenna Interference Source Dataset}
\begin{figure}[]
\centering
\includegraphics[width=3.5in]{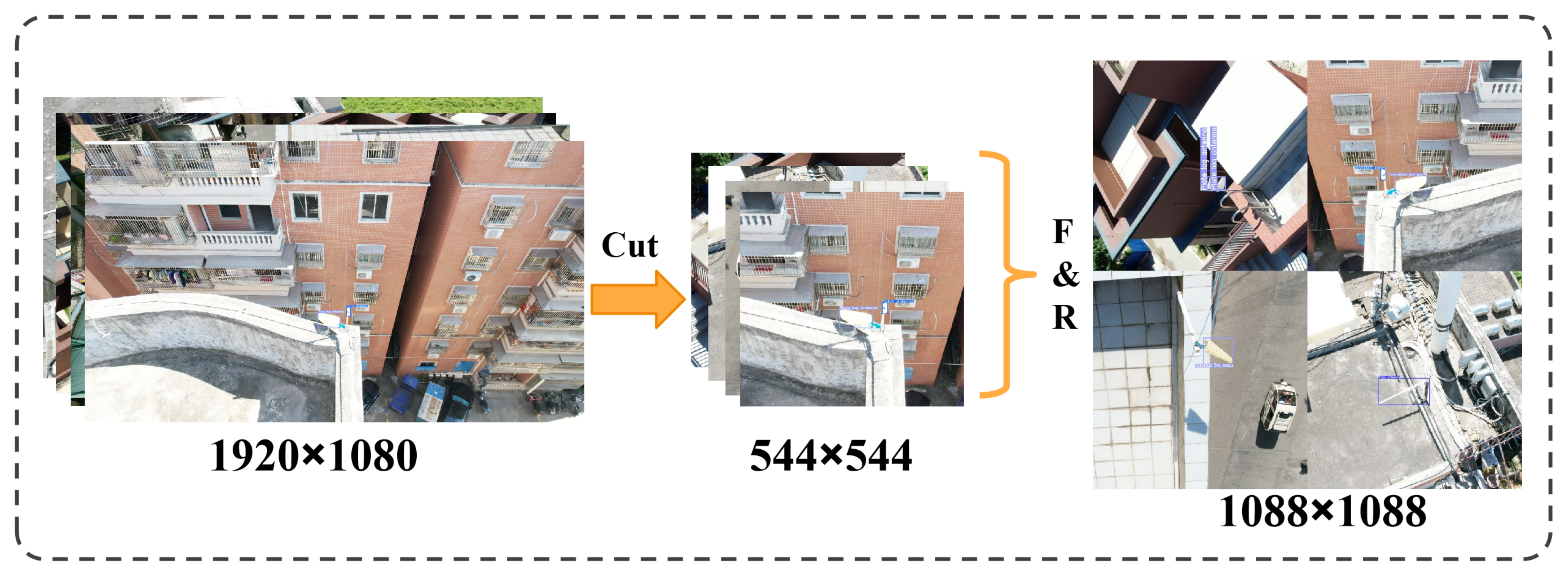}
\caption{Process of constructing the antenna interference source dataset.}
\label{fig10}
\end{figure}

Regarding the antenna interference source inspection task, few existing methods and datasets are available\cite{ref52}. To address this issue, we collaborated with communication professionals to create a dataset by using industrial UAVs that obtained aerial photography for detector analysis and training tasks. We selected three significant antenna interference sources, Yagi, plate logs, and patch antennas, on the basis of routine investigations performed by the communication bureau. Given that single small targets are often contained within one aerial image, we employed an image stitching strategy for dataset augmentation purposes. This involved cropping four randomly selected high-resolution $1920\times1080$ images to obtain $544\times544$ local images containing antenna targets. These local images were then randomly flipped, inverted, and combined to create $1088\times1088$ stitched images, minimizing background interference while preserving small-target characteristics, as shown in Fig. 10. Finally, the dataset was divided into a training set (600) and a validation set (200) at a ratio of 3:1, and professionals labeled it. The entire dataset contained approximately 3200 antenna targets. 

\subsubsection{COCO2017\cite{ref53}}
Microsoft funded and annotated this publicly available dataset in 2014, making it the most authoritative benchmark in the object detection field. Each image contains two to three times more objects than those of other datasets, and COCO2017 is widely used for object detection, image segmentation, and classification tasks. The dataset includes 12 major object categories, 80 subcategories, and 118,278 images.

\textbf{Experimental Configuration}
\setcounter{subsubsection}{0}
\subsubsection{Cloud Server}
We used an Intel Core i9-11900K CPU, an Ubuntu 20.4 system, and an NVIDIA RTX 3060 graphics card to build a modeling approach based on CUDA 11.4 and the PyTorch framework. All detector models were trained on a cloud server. In the experiments, we uniformly set the batch size to 8, the momentum to 0.937, the momentum decay coefficient to 0.0005, and the initial learning rate to 0.01. On the Antenna Interference Source Dataset, 100 training iterations were conducted for Jetson Xavier NX and Raspberry Pi 4B with input sizes 1088×1088 and 640×640, respectively. The COCO dataset was used for 300 epochs of training with an input size of $640\times640$. All model training is conducted without using pre-trained weights.
\subsubsection{Edge Computing Device}
We selected the Jetson Xavier NX from the Jetson series as the UAV's edge computing device. It was equipped with a 384-core NVIDIA Volta GPU with 48 Tensor cores, enabling supercomputer performance at the edge. The power consumption was set at 10W. Additionally, we deployed validation on the Raspberry Pi 4B, which lacked acceleration capabilities. We selected 2GB of RAM and achieved inference using a 64-bit quad-core CPU running at 1.5GHz. Full load power consumption is 7.6W.
\subsubsection{UAV}
We selected the Q600 UAV platform, which was developed by Shenzhen Superway Intelligent Information Technology Co., Ltd. The Q600 flight control board is a Controller Area Network Power Distribution Board (CAN PBD) that utilizes the PX4 open-source solution. The UAV is equipped with an 11000mAh 6s solid-state battery, a flight power consumption of 30W, and a maximum flight speed of up to 3m/s.
\subsubsection{Communication Method}
The RM500U-CN, a 5G Sub-6 GHz module explicitly designed for IoT/eMBB applications, achieves edge-to-cloud communication. It supports an uplink bandwidth of 575 Mbps. In ECC/ECC+ mode, MQTT v5.0 is chosen as the communication protocol. The typical power consumption in idle mode is 0.263W. Quality of Service (QoS) set to 1. In cloud mode, video streaming is processed using the nvvidconv plugin from the GStreamer library, leveraging NVIDIA GPU's hardware acceleration capabilities.
\subsubsection{Multiple UAVs Inspection}
The inspection area is set to $600m \times 600m$, with a UAV coverage radius and grid size of 15m. The UAV flies at a constant speed of 2m/s. Four randomly distributed base stations, spaced 700m apart, ensure signal coverage across the inspection area. The PSO algorithm runs for 300 iterations, with an inertia weight of 0.5 and both personal and global learning factors set to 1.5.

\textbf{Evaluation Criteria}
\setcounter{subsubsection}{0}
\subsubsection{AP and mAP}
In this paper, AP0.5 refers to the mean average precision computed across all categories when the intersection-over-union (IoU) threshold is 0.5. Moreover, mAP denotes the average AP computed at different IoU thresholds ranging from 0.5 to 0.95 (with steps of 0.05), as shown in (22).

\begin{equation}
\small
\mathrm{mAP}=\frac{1}{\mathrm{~N}_{\mathrm{c}}} \sum_{\mathrm{i}}^{\mathrm{N}_{\mathrm{c}}} \mathrm{AP}_{\mathrm{i}}
\end{equation}

\subsubsection{$GFLOPs$ and $Parameters$}
GFLOPs refer to the number of floating-point operations a model performs in a single forward pass. They measure the computational complexity of the tested model and can be used to compare the computational costs of different detectors.
\subsubsection{$FPS$}
This metric represents the average number of frames that can be processed (inferred) within 1 second and is used to evaluate the processing speed of the tested detector.
\subsubsection{$E2EL$ and $Accuracy$}
To more effectively evaluate the relevant indices of the proposed system, we used a UAV in our application to shoot a 40-second inspection video with a flight speed of 1 m/s, which included 22 antenna targets. E2EL is defined in (1). The programs all run as single-threaded processes, with the sampled results being stable data after a 10-second warm-up period. Accuracy is defined as in (23).
\begin{equation}
\small
\mathrm { Accuracy }=\frac{T P}{T P+F N} \times 100 \%
\end{equation}
where TP refers to the number of correct interference source target frames uploaded to the server and FN refers to the number of noninterference source target frames uploaded.
\subsubsection{$Power$ and $RAM$ $Usage$}
Power is used to evaluate the maximum operating time for UAV inspection tasks, excluding the power consumption of cloud servers. RAM Usage is used to measure the resource utilization of edge devices during the inference process.
\subsubsection{$Uplink$ $Bandwidth$}
In this context, it represents the maximum transmission rate at which edge devices in the network upload data to cloud servers—a lower uplink bandwidth results in lower utilization of network resources.

\subsection{Ablation Experiment}
To verify the effectiveness of LHGNet, HetBiFPN, and EL, designed for antenna target detection, we conducted ablation experiments on the antenna interference source dataset, selecting classic or excellent backbones and necks proposed in recent years for comparison. The head was fixed as the decoupled head, which is subsequently expressed as Detect. The experimental results are shown in Table II. We highlight how EdgeAnt achieves a favorable balance between accuracy and computational complexity, which is particularly important for edge devices with limited resources.

\begin{table}[]
\centering
\huge
\renewcommand{\arraystretch}{1.1}
\caption{ABLATION EXPERIMENT CONCERNING THE EDGEANT NETWORK STRUCTURE}
\resizebox{\columnwidth}{!}{%
\begin{tabular}{ccccc}
\hline
Model & AP0.5 & mAP & Params (M) & GFLOPs \\ \hline
\multicolumn{5}{c}{Backbone} \\ \hline
\begin{tabular}[c]{@{}c@{}}CSPDarkNet\cite{ref54}-PANet\cite{ref40}-Detect\\ (YOLOv10-n\cite{ref48})\end{tabular} & 0.697 & 0.401 & 2.6 & 22.2 \\
MobileNetv3\cite{ref55}-v10Neck(PANet\cite{ref40})-Detect & 0.595 & 0.310 & 2.9 & 16.8 \\
EfficientVit\cite{ref56}-v10Neck-Detect & 0.611 & 0.337 & 3.8 & 25.4 \\
ShuffleNetV2\cite{ref57}-v10Neck-Detect & 0.601 & 0.326 & 2.7 & 19.7 \\
RepViTm0.9-v8Neck-Detect\cite{ref58}-v10Neck-Detect & 0.703 & 0.388 & 6.7 & 55.0 \\
ResNet50\cite{ref59}-v10Neck-Detect & 0.632 & 0.335 & \textbf{2.1} & 18.0 \\
HGNetv2\cite{ref39}-v10Neck-Detect & \textbf{0.745} & \textbf{0.439} & 11.2 & 86.3 \\
\rowcolor{gray!20} \textbf{LHGNet-v10Neck-Detect} & 0.724 & 0.409 & 3.6 & \textbf{15.8} \\ \hline
\multicolumn{5}{c}{Neck} \\ \hline
LHGNet-SlimNeck\cite{ref60}-Detect & 0.717 & 0.401 & 2.5 & \textbf{13.5} \\
LHGNet-FPN\cite{ref44}-Detect & 0.692 & 0.390 & 3.5 & 15.0 \\
LHGNet-HSFPN\cite{ref61}-Detect & 0.707 & 0.391 & \textbf{2.8} & 15.6 \\
LHGNet-RepGFPN\cite{ref62}-Detect & 0.716 & 0.403 & 3.2 & 14.0 \\
LHGNet-BiFPN\cite{ref41}-Detect & 0.725 & 0.407 & 3.1 & 14.9 \\
\rowcolor{gray!20} \textbf{LHGNet-HetBiFPN-Detect} & \textbf{0.733} & \textbf{0.419} & 2.9 & 14.5 \\ \hline
\multicolumn{5}{c}{EL} \\ \hline
LHGNet-HetBiFPN-Detect+SPP\cite{ref46} & 0.717 & 0.414 & 3.6 & 15.3 \\
LHGNet-HetBiFPN-Detect+SPPF\cite{ref46} & 0.720 & 0.420 & 3.6 & 15.3 \\
LHGNet-HetBiFPN-Detect+SPPELAN\cite{ref47} & 0.722 & 0.421 & 4.0 & 15.4 \\
LHGNet-HetBiFPN-Detect+PSA\cite{ref48} & 0.722 & \textbf{0.426} & 3.4 & 15.2 \\
\rowcolor{gray!20} \textbf{LHGNet-HetBiFPN-Detect+EL (ours)} & \textbf{0.738} & 0.423 & 3.0 & 14.7 \\ \hline
RTMDet-Tiny\cite{ref63}+EL & \textbf{0.738} & 0.418 & 5.4 & 23.9 \\
YOLOX-Nano\cite{ref64}+EL & 0.613 & 0.311 & \textbf{1.5} & \textbf{6.9} \\
YOLOv8-n+EL & 0.731 & 0.419 & 3.5 & 23.5 \\
YOLOv10-n\cite{ref48}+EL & 0.709 & 0.407 & 2.7 & 22.8 \\ \hline
\end{tabular}%
}
\end{table}

We selected the current SOTA detector (YOLOv10-n)\cite{ref48} for ablation studies. The results showed that utilizing any lightweight backbone network, except for CSPDarkNet\cite{ref54}, significantly decreased the achieved accuracy, as their feature extraction capabilities could not strike a balance between accuracy and a lightweight design. Notably, the computational complexity when using HGNetv2\cite{ref39} as the backbone is 86.3 GFlops, which cannot be run on most low- and mid-end edge devices, such as the Raspberry Pi 4B, which supports a maximum of only 32 GFlops. In contrast, the model using LHGNet has a computational complexity of only one-fifth of that. Although a slight decrease in precision was observed, LHGNet retained most of the advantages of HGNetv2\cite{ref39}, outperforming all existing lightweight backbones.

The significant detector performance improvement achieved with BiFPN\cite{ref41} was due to the scarcity of features for small-target interference sources in large-resolution images, and the bidirectional weighted feature fusion process addresses the feature loss caused by traditional top-down fusion. In addition, after introducing HetConv, the number of model parameters was slightly reduced, and the accuracy was further improved. The $3\times3$ alternating convolutional kernel retained in the HetConv design effectively covered the feature maps of all channels. Moreover, the remaining $1\times1$ convolutional kernel could also aggregate the feature maps to some extent.

Finally, to verify the effectiveness of EL, we conducted experimental validation in 2 dimensions. We first selected a variety of SPPs to add to the connection between the model backbone and the neck. Despite the increased number of parameters, the interference source detection performance significantly declined, and the loss of small-target features due to pyramid pooling was unacceptable. We added EL to several representative single-stage detectors to further demonstrate that EL is adequate. The results showed that the accuracy of each model was improved, and EL could effectively compensate for the feature losses caused by SPPs. The attentional heatmaps produced by various models before and after adding EL are shown in Fig. 11, and the attention paid by the models to small-target objects became more focused to some extent.

\begin{figure}[]
\centering
\includegraphics[width=3.5in]{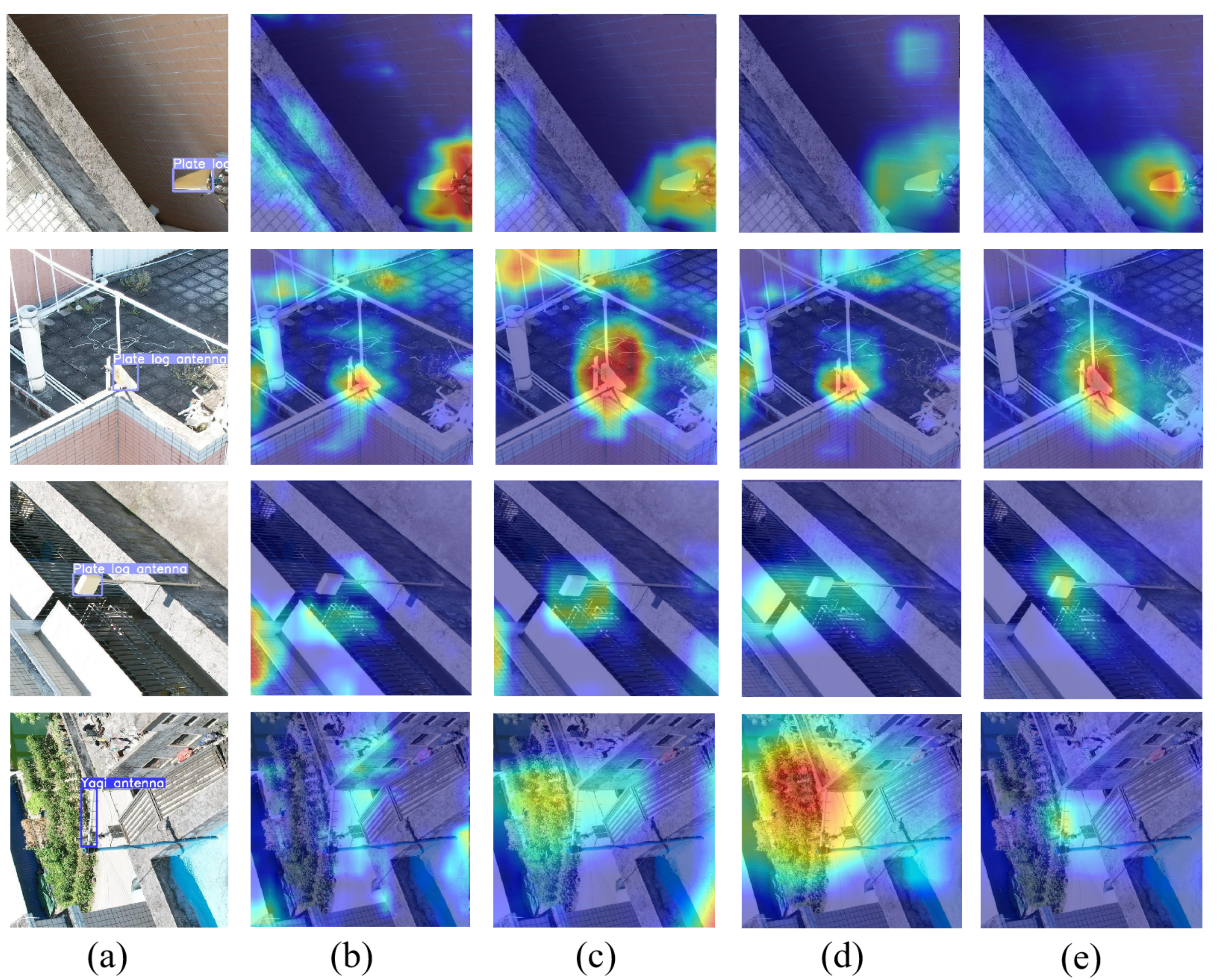}
\caption{Comparison between the heatmaps produced before and after adding EL to different detectors. The four 544x544 subimages were cropped from a 1088x1088 image in the training set. (a) Ground truth. (b) YOLOv8-n. (c) YOLOv8-n+EL. (d) LHGNet-HetBiFPN. (e) LHGNet-HetBiFPN+EL (ours).}
\label{fig11}
\end{figure}

\begin{table}[]
\centering
\caption{ABLATION EXPERIMENT CONCERNING THE LIGHTWEIGHT BACKBONE DESIGN}
\label{tab:my-table}
\huge
\renewcommand{\arraystretch}{1.1}
\resizebox{\columnwidth}{!}{%
\begin{tabular}{cccccccccc}
\hline
\multirow{2}{*}{\#} &
  \multicolumn{2}{c}{HGBlock} &
  \multicolumn{3}{c}{HGNetv2} &
  \multirow{2}{*}{AP0.5} &
  \multirow{2}{*}{mAP} &
  \multirow{2}{*}{Params (M)} &
  \multirow{2}{*}{GFLOPs} \\ \cline{2-6}
  & A          & B          & C          & D          & E          &                         &                         &                        &                         \\ \hline
1 & \multicolumn{5}{c}{Base}                                       & 0.761                   & 0.445                   & 8.8                    & 85.3                    \\
2 & \checkmark &            &            &            &            & 0.769 (+1.1\%)          & 0.453 (+1.8\%)          & 8.9 (+1.1\%)           & 74.9 (-12.2\%)          \\
3 & \checkmark & \checkmark &            &            &            & \textbf{0.785 (+3.2\%)} & \textbf{0.473 (+6.3\%)} & 9.8(+11.4\%)           & 84.2 (-1.3\%)           \\
4 & \checkmark & \checkmark & \checkmark &            &            & 0.733 (-3.7\%)          & 0.428 (-3.8\%)          & 9.8 (+11.4\%)          & 25.7 (-69.9\%)          \\
5 & \checkmark & \checkmark & \checkmark & \checkmark &            & 0.704 (-7.5\%)          & 0.402 (-9.7\%)          & \textbf{2.1 (-76.1\%)} & \textbf{13.0 (-84.8\%)} \\ \rowcolor{gray!20}
6 & \checkmark & \checkmark & \checkmark & \checkmark & \checkmark & 0.738 (-3.0\%)          & 0.423 (-5.0\%)          & 3.0 (-66.0\%)          & 14.7 (-82.8\%)          \\ \hline
\end{tabular}%
}
\end{table}

Table III illustrates the design process from HGNetv2\cite{ref39} to LHGNet, which is central to the lightweight nature of EdgeAnt. "Base" refers to HGNetv2-HetBiFPN-Detect+EL. A and B denote the removal of the excitation convolution from the core backbone component (HGBlock) and the replacement of GhostConv\cite{ref43}, respectively. C, D, and E refer to adjustments made to the HGNetv2\cite{ref39} architecture, specifically involving the addition of transition convolution during the image input process, the trimming of core components, and the replacement of DWConv\cite{ref42} with standard convolution, respectively. LHGNet significantly reduced both the number of parameters and the number of FLOPs compared with those the original HGNetv2\cite{ref39}. Additionally, the loss in model accuracy remained within an acceptable range.

\subsection{Comparative Experiments}
We conducted comparative experiments on both the antenna interference source and COCO datasets, including investigations into detector inference speed, RAM usage, and energy consumption when deployed on edge devices.

\begin{table*}[]
\centering
\caption{PERFORMANCE COMPARISON OF INTERFERENCE SOURCE DETECTION TASKS ON DIFFERENT DEVICES AND WITH OTHER DETECTORS}
\label{tab:my-table}
\Large
\renewcommand{\arraystretch}{1.2}
\resizebox{2.0\columnwidth}{!}{
\begin{tabular}{ccccc|cccc|ccc|cccc}
\hline
\multirow{2}{*}{Model} & \multirow{2}{*}{\begin{tabular}[c]{@{}c@{}}Params\\ (M)\end{tabular}} & \multicolumn{3}{c|}{RTX 3060 (1088)} & \multicolumn{4}{c|}{Jetson Xavier NX (TRT)} & \multicolumn{3}{c|}{RTX 3060 (640)} & \multicolumn{4}{c}{Raspberry Pi 4B (NCNN)} \\ \cline{3-16}
 &  & AP0.5 & mAP & GFLOPs & $\mathrm{mAP}_{1088}^{\mathrm{FP} 16}$ & $\mathrm{FPS}_{1088}^{\mathrm{bs}=1}$ & PCP (W) & GPU (MiB) & AP0.5 & mAP & GFLOPs & $\mathrm{mAP}_{640}^{\mathrm{INT} 8}$ & $\mathrm{FPS}_{640}^{\mathrm{bs}=1}$ & PCP (W) & CPU (MiB)\\ \hline
YOLOv3-MobileNetv2\cite{ref65} & 3.7 & 0.702 & 0.328 & 18.4 & 0.228 & 17.2 & 5.0 & 712 & 0.605 & 0.256 & 6.3 & 0.138 & 2.5 & 2.98 & 824 \\
YOLOv4-tiny\cite{ref64} & 5.9 & 0.681 & 0.384 & 23.4 & 0.317 & 15.3 & 5.5 & 876 & 0.607 & 0.294 & 8.1 & 0.167 & 2.8 & 3.32 & 745 \\
YOLOv5-n & 2.5 & 0.656 & 0.364 & 20.1 & 0.297 & 14.9 & 4.4 & 396 & 0.612 & 0.322 & 7.1 & 0.196 & 4.4 & \textbf{2.63} & 483 \\
YOLOv6-n\cite{ref66} & 4.4 & 0.688 & 0.391 & 33.5 & 0.330 & 12.4 & 5.3 & 817 & 0.629 & 0.336 & 11.8 & 0.220 & 2.6 & 2.68 & 561 \\
YOLOv7-n\cite{ref67} & 3.0 & 0.638 & 0.369 & 27.2 & 0.298 & 12.8 & 5.1 & 629 & 0.609 & 0.325 & 10.5 & 0.145 & 2.7 & 2.75 & 548 \\
YOLOv8-n & 3.0 & 0.717 & 0.403 & 23.0 & 0.345 & 16.5 & 4.8 & 505 & 0.637 & 0.338 & 8.2 & 0.223 & 3.5 & 2.81 & 553 \\
GELAN-t\cite{ref47} & 2.4 & 0.678 & 0.382 & 29.1 & 0.318 & 12.3 & 5.2 & 752 & 0.550 & 0.308 & 10.7 & 0.212 & 2.8 & 3.46 & 847 \\
Yolov10-n\cite{ref48} & 2.6 & 0.697 & 0.401 & 22.2 & 0.341 & 15.8 & 4.6 & 694 & 0.620 & 0.330 & 7.9 & 0.219 & 3.9 & 3.24 & 782 \\
YOLOX-Nano\cite{ref64} & \textbf{1.4} & 0.591 & 0.308 & \textbf{5.3} & 0.196 & 15.5 & 4.3 & \textbf{215} & 0.487 & 0.215 & \textbf{1.9} & 0.098 & 3.7 & 3.06 & \textbf{423} \\
PP-YOLOE-Plus-s\cite{ref68} & 7.5 & 0.718 & 0.415 & 23.0 & \textbf{0.370} & 9.2 & 4.7 & 991 & 0.652 & 0.349 & 7.9 & 0.237 & 2.3 & 3.61 & 951 \\
RTMDet-Tiny\cite{ref63} & 4.9 & 0.694 & 0.402 & 23.2 & 0.334 & 11.9 & 5.1 & 748 & 0.634 & 0.331 & 8.1 & 0.225 & 4.1 & 2.93 & 628 \\
SSD-MobileNetv2\cite{ref69} & 3.0 & 0.652 & 0.367 & 8.0 & 0.296 & 16.7 & 5.4 & 305 & 0.598 & 0.317 & 2.8 & 0.158 & 3.2 & 3.15 & 598 \\
EfficientDet-D0\cite{ref41} & 3.8 & 0.670 & 0.366 & 9.4 & 0.307 & 17.1 & 4.9 & 968 & 0.617 & 0.329 & 3.6 & 0.172 & 4.5 & 2.93 & 654 \\ \hline \rowcolor{gray!20}
EdgeAnt (ours) & 3.0 & \textbf{0.735} & \textbf{0.421} & 14.7 & 0.367 & \textbf{21.1} & \textbf{4.2} & 640 & \textbf{0.688} & \textbf{0.352} & 5.1 & \textbf{0.248} & \textbf{4.8} & 2.67 & 569 \\ \hline
\end{tabular}
}
\end{table*}

\begin{figure}[]
\centering
\includegraphics[width=3.5in]{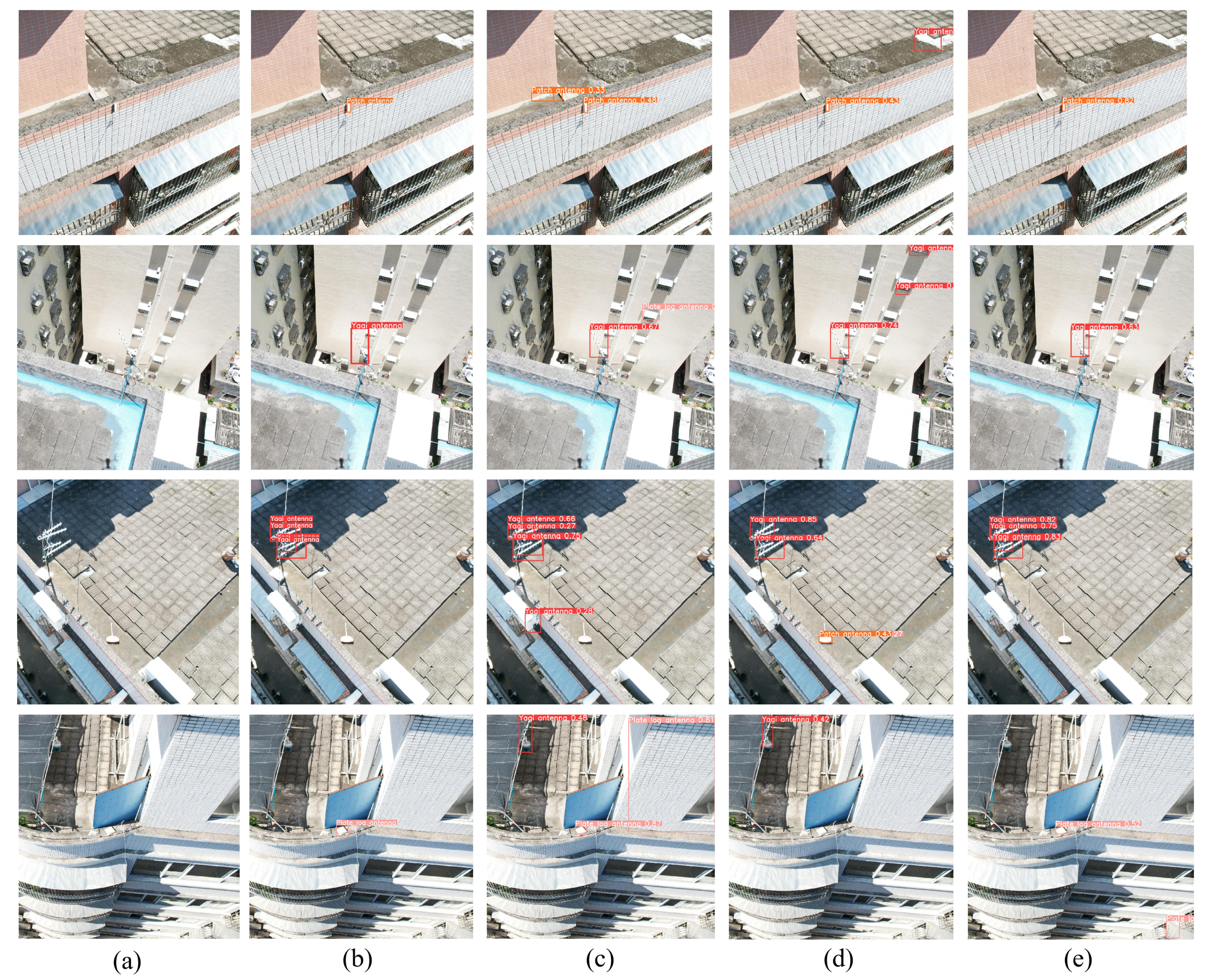}
\caption{Prediction results yielded by different detectors on the interference source dataset. (b) Ground truth. (c) YOLOX-Nano. (d) YOLOv8-n (e) Our method.}
\label{fig12}
\end{figure}

Table IV presents the comparative test results obtained on the antenna interference source dataset. The results show that YOLOv10\cite{ref48} did not outperform YOLOv8 on the interference source detection task as expected, which may have been because YOLOv10\cite{ref48} uses a large $7\times7$ kernel convolution in its compact inverted block (CIB). This approach is not conducive to tiny interference source targets, and part of the enlarged sensory field is redundant, as reflected in the design of TSRBlock. EdgeAnt employs a feature map channel expansion strategy through standard convolution during the image input phase, which significantly reduces the number of GFLOPs and accelerates the inference speed of the model; this approach initially causes the model to lose a certain number of practical features. The subsequent hierarchical feature enhancement and bidirectional feature fusion steps effectively compensate for this deficiency.

We deployed the model on two edge devices, Jetson Xavier NX, and Raspberry Pi 4B, using TensorRT (FP16) and NCNN (INT8) for inference acceleration. Considering the Raspberry Pi 4B lacks a GPU for acceleration, we adjusted the input size to 640×640. It can be observed that models with higher computational complexity are typically more constrained in inference speed on edge devices. With the support of TensorRT, EdgeAnt's lightweight advantage in terms of GFLOPs became more pronounced, achieving an FPS of 22.1, making it the only detector surpassing 20 FPS. Correspondingly, the power consumption of processes on edge computing devices was the lowest. The inevitable precision loss due to model quantization was observed, especially evident in models with fewer parameters like YOLOX-Nano. The results indicate that EdgeAnt effectively overcame the limitations of limited resources.

\begin{table}[]
\centering
\caption{COMPARISON WITH OTHER DETECTORS ON THE COCO DATASET}

\renewcommand{\arraystretch}{1.1}
\label{tab:my-table}
\resizebox{\columnwidth}{!}{%
\begin{tabular}{ccccc}
\hline
Model & AP0.5 & mAP & \begin{tabular}[c]{@{}c@{}}Params (M)\end{tabular} & \begin{tabular}[c]{@{}c@{}}GFLOPs\end{tabular} \\ \hline
YOLOv3-MobileNetv2\cite{ref65}         & 0.374          & 0.254          & 3.7          & 6.5          \\
YOLOv4-Tiny\cite{ref70} & 0.421          & 0.249          & 6.1          & 8.2          \\
YOLOv6-n\cite{ref66}                   & 0.476          & 0.332          & 4.5          & 13.0         \\
YOLOv7-Tiny\cite{ref67}                & 0.528          & 0.352          & 6.2          & 6.9          \\
YOLOv8-n                                & 0.518          & 0.369          & 3.2          & 8.7          \\
YOLOv9-t\cite{ref47}                  & 0.525          & 0.379          & 3.7 & 16.2         \\
YOLOv10-n\cite{ref48}                 & 0.530          & 0.379          & \textbf{2.8}          & 8.5          \\
YOLOX-Tiny\cite{ref64}                  & 0.503          & 0.328          & 5.1          & 7.6          \\
PP-YOLOE-Plus-s\cite{ref68}           & \textbf{0.602} & \textbf{0.435} & 7.9          & 8.7          \\
RTMDet-Tiny\cite{ref63}               & 0.569          & 0.403          & 4.9          & 8.1          \\ \hline \rowcolor{gray!20}
EdgeAnt (ours)                          & 0.531          & 0.389          & 3.2 & \textbf{5.3} \\ \hline
\end{tabular}%
}
\end{table}

\begin{figure}[]
\centering
\includegraphics[width=3.5in, height=2.3in]{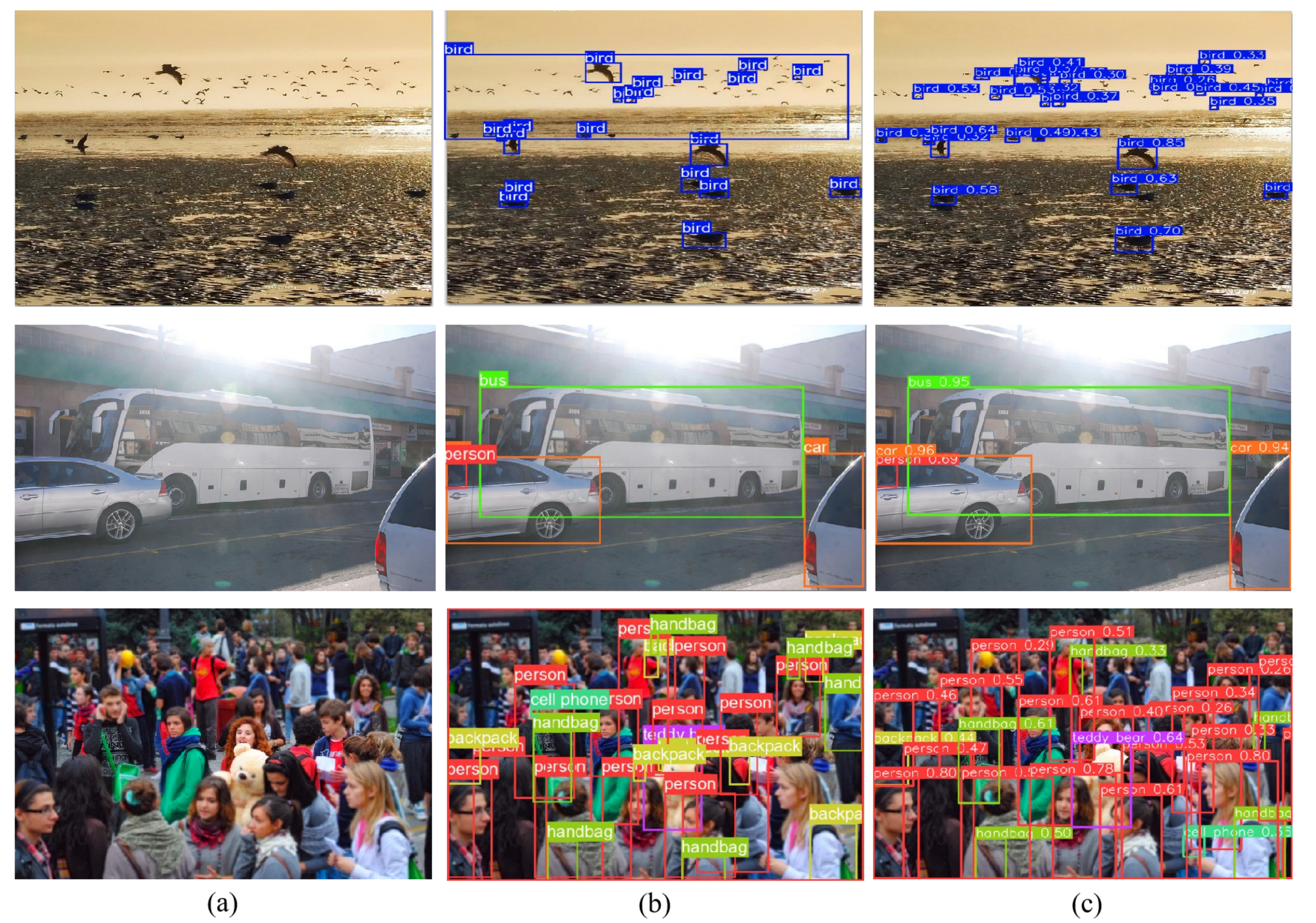}
\caption{Performance achieved by EdgeAnt in terms of reasoning about complex background images and images with intense lighting. (b) Ground truth. (c) Our method.}
\label{fig13}
\end{figure}

The visual inference results produced by different detectors are shown in Fig. 12. Under bright lights, roof walls and clutter exhibit remarkably similar characteristics to those of plate logs and patch antennas, but EdgeAnt could still distinguish them well. As shown in the third picture, EdgeAnt could accurately distinguish each Yagi antenna in the case with occlusion. EdgeAnt proved to be robust enough to adapt to detecting interference sources in different scenarios. Notably, YOLOX-Nano\cite{ref64} mistakenly detected the massive wall as an interference source in the fourth figure, proving that the KSA is meaningful.

To verify the generalizability of EdgeAnt, we conducted experiments on the COCO dataset; the results are shown in Table IV. Compared with YOLOv8-n possessing the same number of parameters, EdgeAnt yielded a 5.4\% higher mAP and required 61\% fewer FLOPs, which are very competitive results. The results show that the network architecture of EdgeAnt is capable of capturing and adapting to a wide range of target features. For mobile platforms, accurately extracting the feature information of targets located in complex images while keeping the model lightweight is crucial. We selected several representative images with cluttered backgrounds and strong lighting conditions from the COCO dataset, and the inference results produced by EdgeAnt are shown in Fig. 13. As shown in the figure, EdgeAnt could accurately detect different objects, especially small-target scenes at long distances. The EL layer between the EdgeAnt backbone and neck functioned as expected, fully enhancing the expression of small-target features through attention mechanisms.

In the ECC+ mode, the efficient lightweight interference source localization algorithm enhanced system latency, whereas the KSA optimized operational efficiency. Fig. 14 illustrates how we determined the KSA parameters. Given that the overhead view angle of the UAV was relatively constant, the sizes of the interference sources in a single image remained stable. Among the three types of interference sources, the panel antenna was most prone to false detections, making it essential to apply a pixel threshold to filter the detector outputs. The Yagi antenna, which was the largest type, occupied 100-120 pixels, so we set $\tau$ to 120. Regarding the tracking threshold $\mu$, a low threshold led to many false targets being uploaded. The light blue interval shows the effective range for edge devices to upload the detected targets. However, setting $\mu$ too high may result in missed detections, so we set it to 6.

\subsection{System Validation}
\begin{figure}[]
\centering
\includegraphics[width=3.5in,height=1.3in]{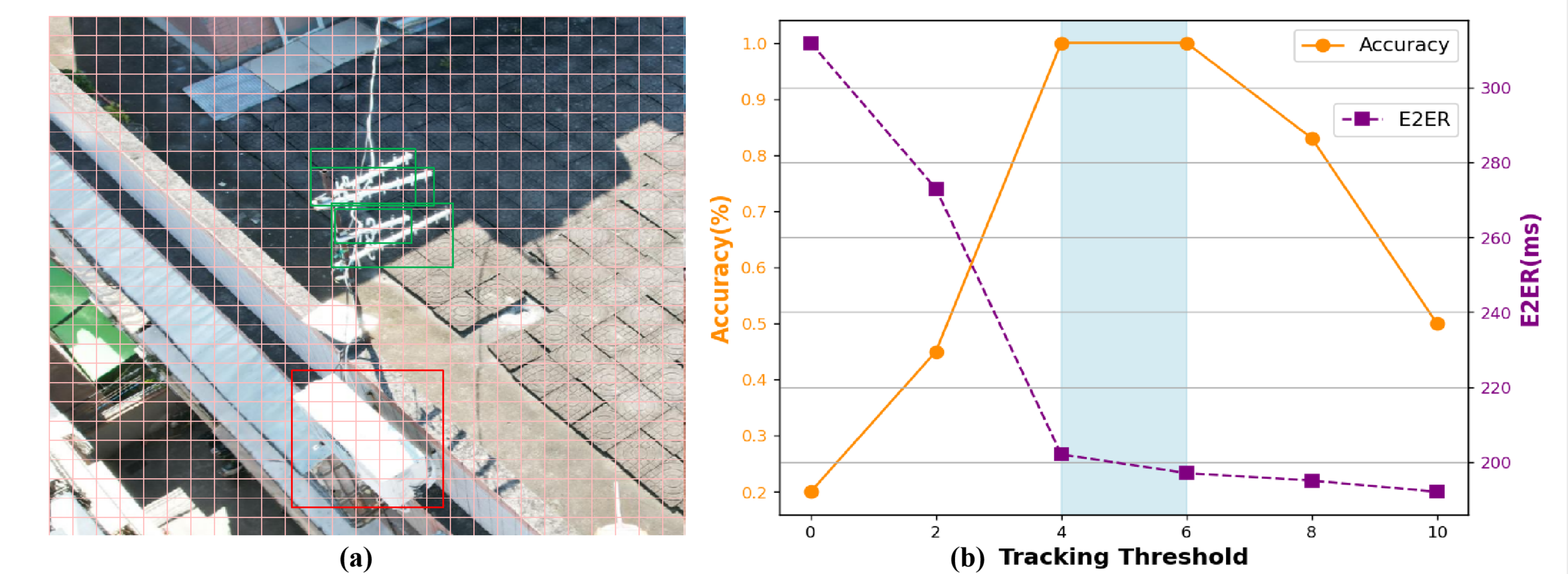}
\caption{Critical parameter selection for the KSA. (a) presents a 1080 $\times$ 1080 image segmented by horizontal and vertical reference lines spaced at 20 pixels. The correctly identified interference source is marked with a green box, whereas incorrect targets are marked with red. (b) presents the test results obtained with tracking thresholds ranging from 1-10 in the simulated videos.}
\label{fig14}
\end{figure}

\begin{table}[]
\centering
\caption{SYSTEM PERFORMANCE EVALUATION UNDER DIFFERENT INSPECTION ALGORITHMS AND APPLICATION MODES}
\label{tab:my-table}
\renewcommand{\arraystretch}{1.1}
\Large
\resizebox{\columnwidth}{!}{%
\begin{tabular}{c|ccccc}
\hline
Mode &
  Method &
  \begin{tabular}[c]{@{}c@{}}E2EL\\ (ms)\end{tabular} &
  \begin{tabular}[c]{@{}c@{}}Accuracy \\ (\%)\end{tabular} &
  \begin{tabular}[c]{@{}c@{}}Power\\ (W)\end{tabular} &
  \begin{tabular}[c]{@{}c@{}}GPU\_{\text{NX}}\\ (MiB)\end{tabular} \\ \hline
\multirow{3}{*}{CO}  & YOLOv8-n+BotSort       & 1012(C)+14.1(I)        & 75.0          & 3.1(C)+9.8(I)           & \textbf{64.1} \\
                     & YOLOv8-n+AntSort       & 1012(C)+12.4(I)        & 79.2          & 3.1(C)+9.8(I)           & \textbf{64.1} \\
                     & \cellcolor{gray!20}EdgeAnt+AntSort  & \cellcolor{gray!20}1012(C)+7.9(I)         & \cellcolor{gray!20}86.4          & \cellcolor{gray!20}3.1(C)+9.8(I)           & \cellcolor{gray!20}\textbf{64.1} \\ \hline
\multirow{3}{*}{ECC} & YOLOv8-n+BotSort       & 251(C)+82.6(I)         & 75.0          & 1.4(C)+13.9(I)          & 514.2         \\
                     & YOLOv8-n+AntSort       & 251(C)+79.5(I)         & 79.2          & 1.4(C)+13.7(I)          & 513.9         \\
                     & \cellcolor{gray!20}EdgeAnt+AntSort & \cellcolor{gray!20}251(C)+51.3(I)         & \cellcolor{gray!20}86.4          & \cellcolor{gray!20}1.4(C)+12.1(I)          & \cellcolor{gray!20}651.8         \\ \hline
\multirow{3}{*}{\begin{tabular}[c]{@{}c@{}}ECC+\\ (ours)\end{tabular}} &
  YOLOv8-n+BotSort &
  62(C)+82.6(I) &
  78.2 &
  0.3(C)+13.8(I) &
  503.8 \\
                     & YOLOv8-n+AntSort       & 62(C)+79.5(I)          & 82.6          & 0.3(C)+13.5(I)          & 501.2         \\
                     & \cellcolor{gray!20}EdgeAnt+AntSort & \cellcolor{gray!20}\textbf{62(C)+51.3(I)} & \cellcolor{gray!20}\textbf{90.4} & \cellcolor{gray!20}\textbf{0.3(C)+12.1(I)} & \cellcolor{gray!20}645.4         \\ \hline
\end{tabular}%
}
\end{table}

\begin{table}[]
\centering
\caption{SYSTEM PERFORMANCE EVALUATION UNDER DIFFERENT RESOLUTION VIDEO STREAMS AND APPLICATION MODES}
\label{tab:my-table}
\renewcommand{\arraystretch}{1.1}
\resizebox{\columnwidth}{!}{%
\begin{tabular}{c|ccccc}
\hline
Mode &
  \begin{tabular}[c]{@{}c@{}}Input\\ Video\end{tabular}  &
  \begin{tabular}[c]{@{}c@{}}E2EL\\ (ms)\end{tabular} &
  \begin{tabular}[c]{@{}c@{}}Accuracy \\ (\%)\end{tabular} &
  \begin{tabular}[c]{@{}c@{}}Power\\ (W)\end{tabular} &
  \begin{tabular}[c]{@{}c@{}}GPU$_\text{NX}$\\ (MiB)\end{tabular} \\ \hline
\multirow{3}{*}{CO}  & 480P(120f) & 398(C)+3.1(I)  & 32.9          & 2.4(C)+9.6(I)  & \textbf{22.3} \\
                     & 720P(60f)  & 728(C)+4.4(I)  & 68.2          & 2.6(C)+9.6(I)  & 36.6          \\
                     & \cellcolor{gray!20}1080P(30f) & \cellcolor{gray!20}1012(C)+7.9(I) & \cellcolor{gray!20}86.4          & \cellcolor{gray!20}3.1(C)+9.8(I)  & \cellcolor{gray!20}64.1          \\ \hline
\multirow{3}{*}{ECC} & 480P(120f) & 243(C)+21.5(I) & 32.9          & 0.9(C)+10.7(I) & 583.8         \\
                     & 720P(60f)  & 235(C)+31.1(I) & 68.2          & 1.1(C)+11.0(I) & 615.1         \\
                     & \cellcolor{gray!20}1080P(30f) & \cellcolor{gray!20}251(C)+51.3(I) & \cellcolor{gray!20}86.4          & \cellcolor{gray!20}1.4(C)+12.1(I) & \cellcolor{gray!20}651.8         \\ \hline
\multirow{3}{*}{\begin{tabular}[c]{@{}c@{}}ECC+\\ (ours)\end{tabular}} &
  480P(120f) &
  \textbf{51(C)+21.5(I)} &
  37.8 &
  \textbf{0.3(C)+10.7(I)} &
  576.8 \\
                     & 720P(60f)  & 54(C)+31.1(I)  & 74.1          & 0.3(C)+11.0(I) & 599.5         \\
                     & \cellcolor{gray!20}1080P(30f) & \cellcolor{gray!20}62(C)+51.3(I)  & \cellcolor{gray!20}\textbf{90.4} & \cellcolor{gray!20}0.4(C)+12.5(I) & \cellcolor{gray!20}621.4         \\ \hline
\end{tabular}%
}
\end{table}

We simulated actual UAV flight attitudes using IMU data collected from inspection videos and compared the system performance of different detector and tracker combinations across various application modes. "C" and "I" represent the communication and inference components. In CO mode, the edge device is responsible only for video capture and transmission, sending detection requests to the cloud server via the SDK and using the RTSP protocol for video stream transmission. Considering the bandwidth limitations in real-world scenarios, we limited the uplink bandwidth to 40 Mb/s. Table VI shows that despite GPU-optimized video encoding in CO mode, real-time video streaming still results in significant E2EL. It imposes a load on the 5G communication module. The ECC mode begins using the MQTT protocol to upload inference results, alleviating the network load, but a substantial amount of redundant data is still uploaded. In ECC+ mode, only keyframes are uploaded, optimizing the E2EL to within 150 ms. Compared with the YOLOv8-n and BotSort\cite{ref51} combination in CO mode, our method reduces E2EL by 88.9\%. Notably, CO and ECC modes exhibit higher false detection rates due to the lack of KSA filtering for detection and tracking results.

\begin{figure}[]
\centering
\includegraphics[width=3.5in]{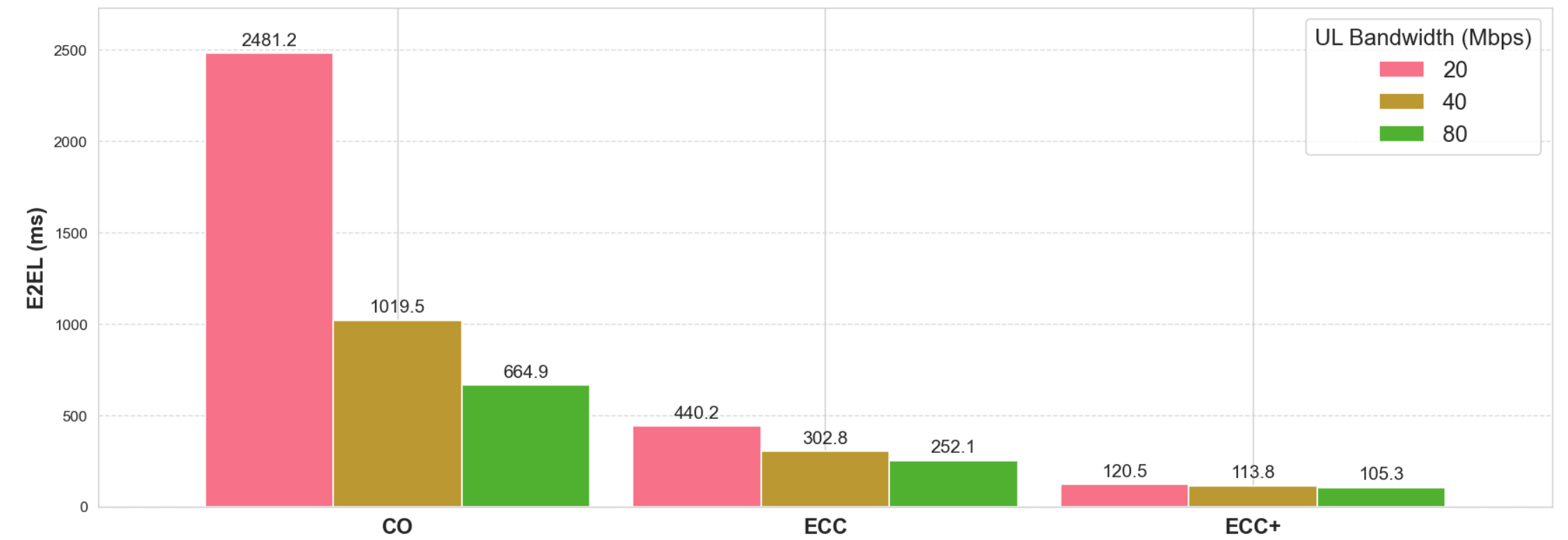}
\caption{Performance comparison of different system modes under variable upstream bandwidth. The selected inspection algorithm combination is "EdgeAnt+AntSort."}
\label{fig15}
\end{figure}

To thoroughly validate the robustness of our system, we conducted further system performance tests based on the "EdgeAnt+BoTSort" framework using video streams of different resolutions as input. The experimental results are shown in Table VII. Low-resolution images lack the antenna's texture information, making them unable to meet the accuracy requirements of actual inspections. In CO mode, video transmission latency exponentially grows as the resolution increases. Additionally, we compared the system performance under varying bandwidth scenarios with the experimental results shown in Fig. 15. It can be observed that from CO to ECC+ mode, the influence of video resolution and bandwidth on system performance gradually diminishes, shifting the primary determinant of system performance to the model itself.

\begin{figure}[]
\centering
\includegraphics[width=3.5in]{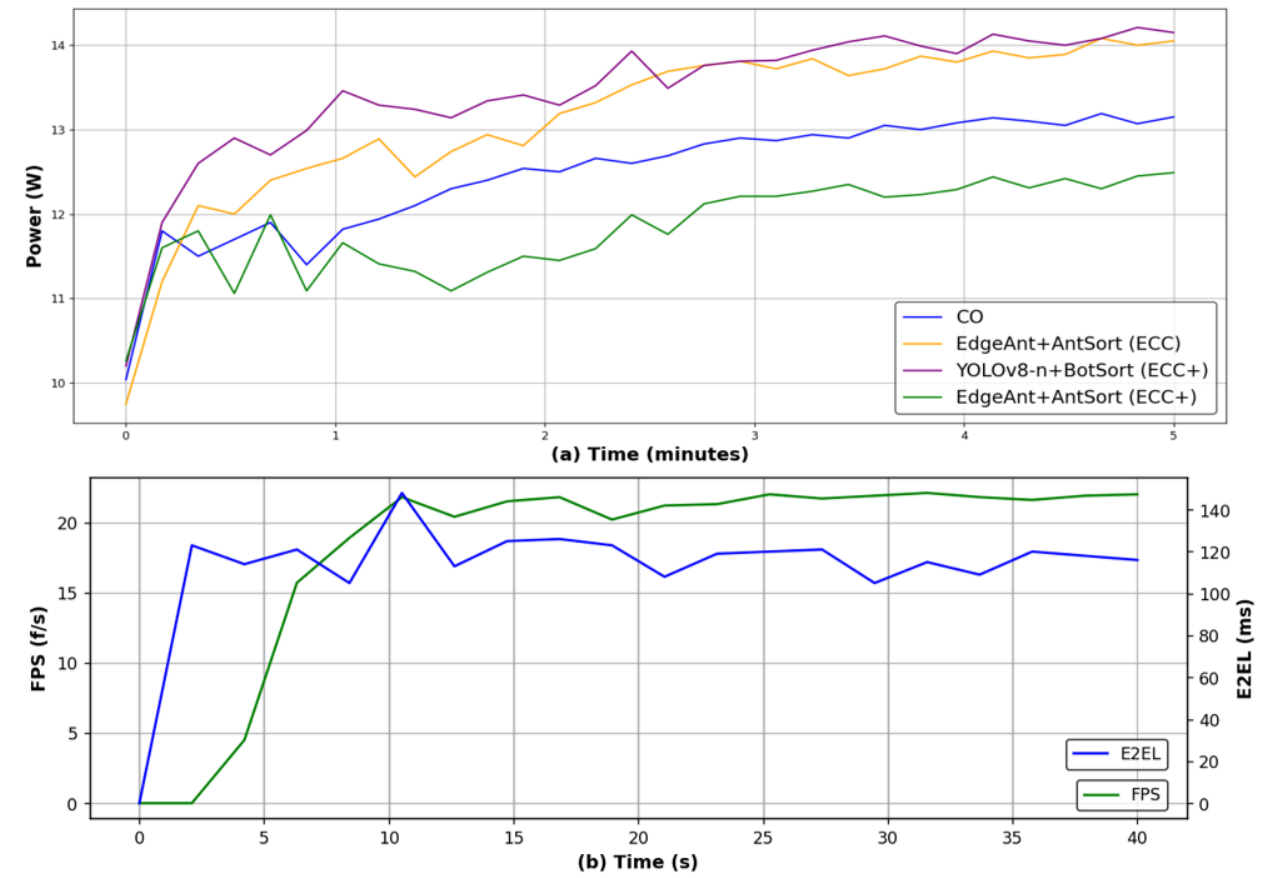}
\caption{System stability validation. (a) shows the power variations exhibited by Jetson Xavier NX and 5G mobile module for different combinations of AI modes and localization algorithms, where the test time was 5 minutes. (b) shows the change exhibited by the E2EL of our system for the test video.}
\label{fig16}
\end{figure}

In Fig. 16(a), we illustrate the overall energy consumption levels of edge computing devices and 5G mobile module across various critical configurations. In the CO mode, despite the absence of AI inference tasks on the edge device, the continuous hardware-accelerated encoding and streaming of real-time video streams exert pressure on the CPU and GPU, affecting system stability. Conversely, in the ECC+ system mode, the 5G mobile module remains low-power for extended periods, resulting in minimal transmission energy consumption. Fig. 16(b) shows the stability of the localization algorithm, which maintained an inference FPS above 20 after initialization. E2EL peaked at around 10 seconds due to concentrated interference source uploads, with fluctuations under 30 milliseconds in other cases.

\begin{figure}[]
\centering
\includegraphics[width=3.5in]{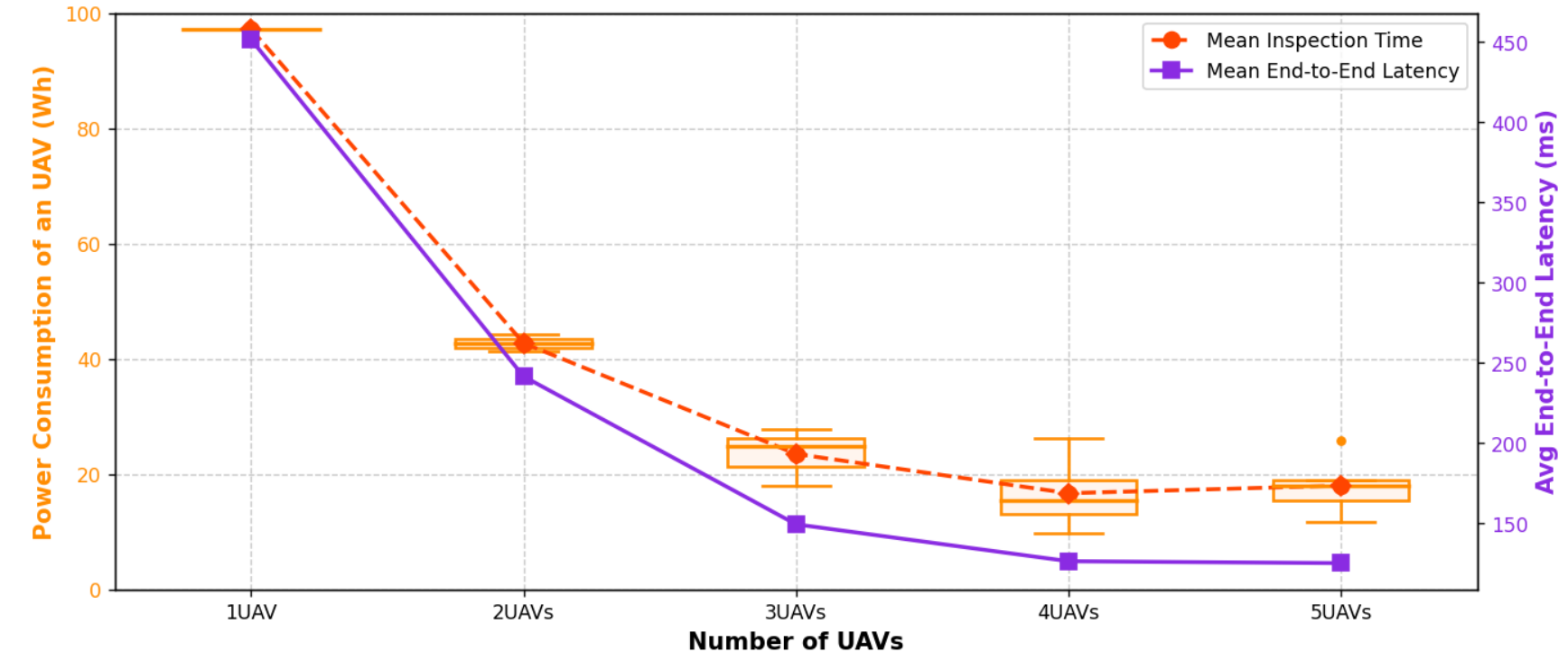}
\caption{Comparison of inspection efficiency with different numbers of UAVs.}
\label{fig17}
\end{figure}

We simulated multiple UAVs performing cooperative inspection based on the objective function defined in (2), as shown in Fig. 17. A single UAV’s battery capacity is insufficient to cover the 360,000 $m^2$ inspection area. As the number of deployed UAVs increases, each UAV has more opportunities to select routes closer to the base station, reducing delay. However, when the number of UAVs reaches five, airspace congestion increases average energy consumption. This demonstrates that proper deployment and path planning can improve inspection efficiency in practical applications.

\section{Conclusion}
This paper designs an AIoT system for UAV antenna interference source inspection based on the ECC+ mode. It includes a complete TBD interference source localization scheme consisting of a novel lightweight detector (EdgeAnt) and an improved tracker (AntSort), which achieve high-precision and rapid inference on resource-constrained edge devices. KSA selectively uploads inspection results compared to ECC mode, optimizing system latency and accuracy. We conducted a comprehensive performance evaluation, showing that EdgeAnt is the best interference detector. With the assistance of AntSort, our system achieved an 88.9\% reduction in E2EL compared to the CO mode. The system also demonstrated robustness and stability across input resolutions and bandwidths. Additionally, it exhibits good scalability for multiple UAVs inspections in practical applications.

In the future, we aim to develop a more universally adaptable TBD architecture to accommodate inspection tasks across various scenarios and devices. This will include a universal model with subdecimal parameters and a tracker capable of handling complex motion dynamics and target occlusions. In addition, further research is warranted on optimizing scan coverage and cooperative complementarity in dynamic multiple UAVs inspection environments while considering communication interference among UAVs.

\bibliographystyle{ieeetr}
\bibliography{ref}
\vspace{-25pt}

\begin{IEEEbiography}[{\includegraphics[width=1in,height=1.25in,clip,keepaspectratio]{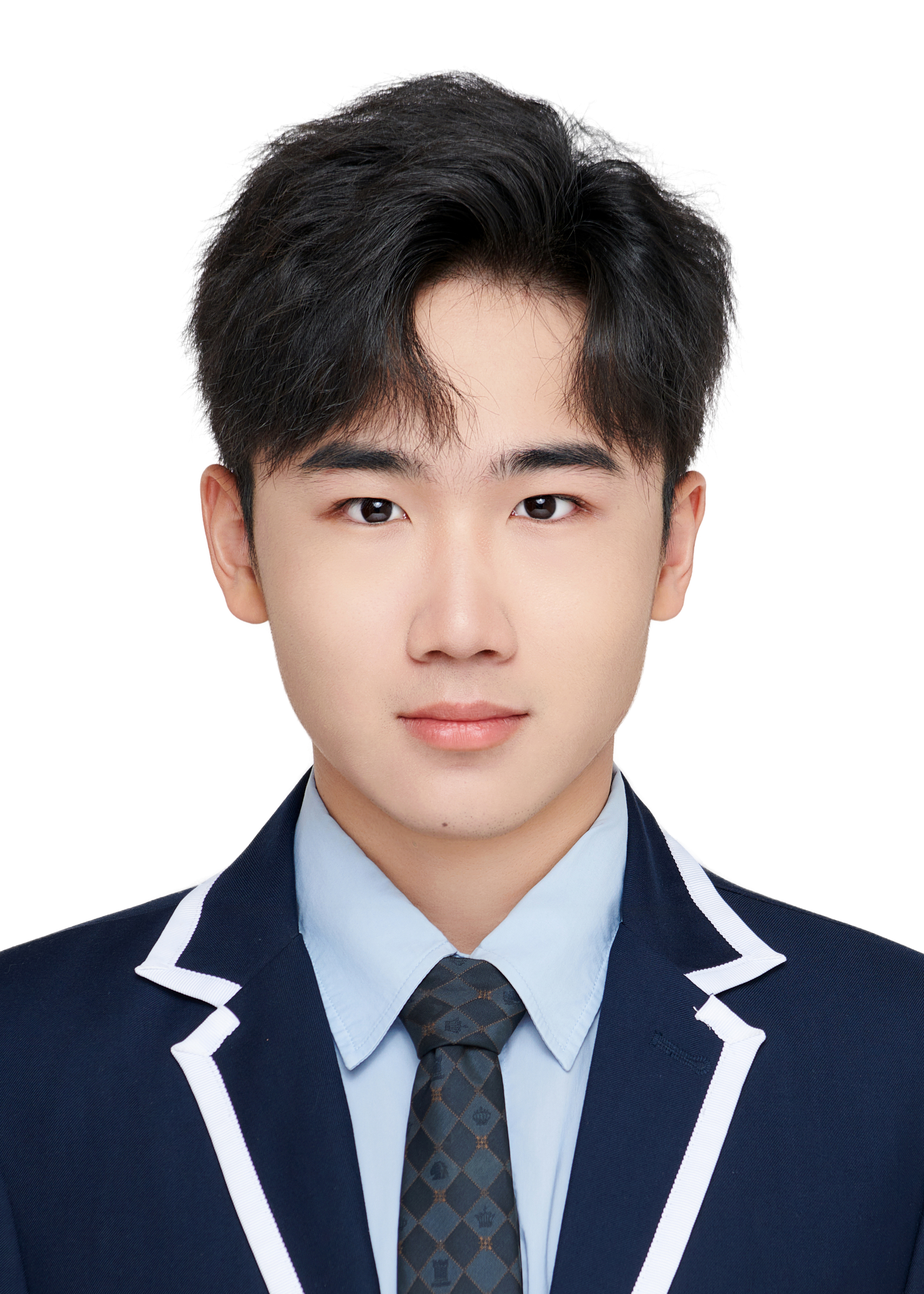}}]
{Jun Dong} (Student Member, IEEE) is currently pursuing the bachelor’s degree in Internet of Things (IoT) engineering with the School of Data Science and Engineering, South China Normal University, Shanwei, China. 

His main research interests include embedded development, artificial intelligence, and target detection.
\end{IEEEbiography}

\vspace{-25pt}
\begin{IEEEbiography}[{\includegraphics[width=1in,height=1.25in,clip,keepaspectratio]{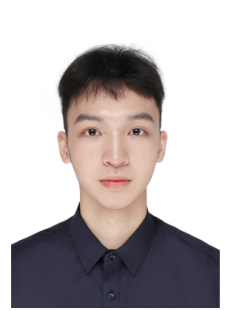}}]{Jintao Cheng}
received his bachelor s degree from the School of Physics and Telecommunications Engineering, South China Normal University, in 2021. His research focuses on computer vision, SLAM and deep learning.
\end{IEEEbiography}

\vspace{-25pt}
\begin{IEEEbiography}[{\includegraphics[width=1in,height=1.25in,clip,keepaspectratio]{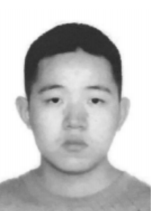}}]{Jin Wu}
(Member, IEEE) was born in Zhenjiang, China, in 1994. He received a B.S. degree from the University of Electronic Science and Technology of China, Chengdu, China. He is currently pursuing a Ph.D. degree with the Robotics and Multiperception Lab, Hong Kong University of Science and Technology (HKUST), Hong Kong, under the supervision of Prof. M. Liu. He has been a research assistant with the Department of Electronic and Computer Engineering, HKUST, since 2018. He has coauthored over 50 technical papers in representative journals and conference proceedings of IEEE, AIAA, and IET. One of his papers published in IEEE TRANSACTIONS ON AUTOMATION SCIENCEAND ENGINEERING was selected as a ESI Highly Cited Paper by the ISI Web of Science from 2017-2018. His research interests include robot navigation, multisensor fusion, mechatronics, and robotic application circuitization.
\end{IEEEbiography}
\vspace{-25pt}

\begin{IEEEbiography}[{\includegraphics[width=1in,height=1.25in,clip,keepaspectratio]{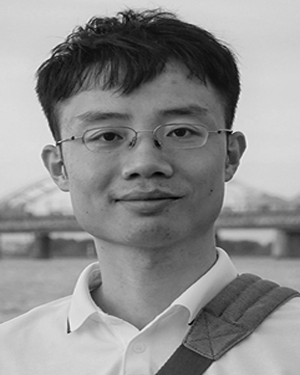}}]{Chengxi Zhang}
received B.S. and M.S. degrees from the Harbin Institute of Technology, Harbin, China, in 2012 and 2015, respectively, and a Ph.D. degree from Shanghai Jiao Tong University, Shanghai, China, in 2019.

He is currently an associate professor with Jiangnan University, Wuxi, China. His interests include robotics and control.

Dr. Zhang is a Session Chair of JACA2022, Wuxi, and an Invited Session Chair of the 36th CCDC2024, Xi an. He is an Associate Editor of Frontiers in Aerospace Engineering and an Editorial Board Member of IoT, Applied Math, AI and Autonomous Systems, Aerospace Systems, and Astrodynamics.
\end{IEEEbiography}
\vspace{-25pt}

\begin{IEEEbiography}[{\includegraphics[width=1in,height=1.25in,clip,keepaspectratio]{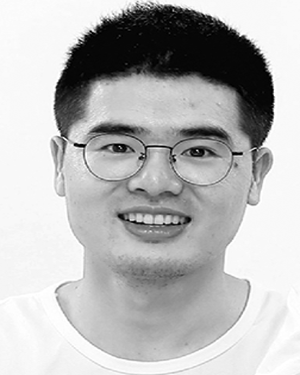}}]{Xiaoyu Tang}
(Senior Member, IEEE) received a Ph.D. degree in control theory and applications from the Key Laboratory of Advanced Process Control for the Light Industry (Ministry of Education), Institute of Automation, Jiangnan University, Wuxi, China, in 2015.

From 2013-2014, he was a visiting student with the Department of Chemical and Materials Engineering, University of Alberta, Edmonton, AB, Canada, where he was a postdoctoral fellow from 2015-2018. In 2015, he joined Jiangnan University
as an associate professor, where he is currently a professor. His research interests include statistical signal processing, Bayesian estimation theory, and fault detection and diagnosis.

Dr. Zhao was a recipient of the Alexander von Humboldt Research Fellowship in Germany and the Excellent Ph.D. Thesis Award in Jiangsu Province, China, in 2016.
\end{IEEEbiography}
\vspace{-25pt}

\begin{IEEEbiography}[{\includegraphics[width=1in,height=1.25in,clip,keepaspectratio]{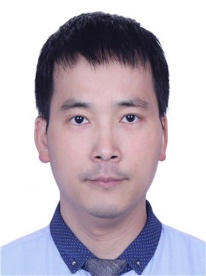}}]{Xiaoyu Tang} (Member, IEEE) received the B.S. degree from South China Normal University, Shanwei, China, in 2003, and the M.S. degree from Sun Yat-sen University, Guangzhou, China, in 2011. 

He is currently pursuing the Ph.D. degree with South China Normal University. He is working with Xingzhi College, South China Normal University, where he is engaged in information system development. His research interests include machine vision, intelligent control, and the Internet of Things.

Mr. Tang is a member of the IEEE ICICSP Technical Committee.
\end{IEEEbiography}

\end{document}